\documentstyle[12pt]{article}

\newcommand{\beq}{\begin{eqnarray}}
\newcommand{\eeq}{\end{eqnarray}}
\makeatletter
\def\vereq#1#2{\lower3pt\vbox{\baselineskip1.5pt \lineskip1.5pt
\ialign{$\m@th#1\hfill##\hfil$\crcr#2\crcr\sim\crcr}}}
\makeatother

\begin{document}

\begin{titlepage}

\begin{center}
\vspace*{-1cm}
SCIPP-99-49    \hfill    NSF-ITP-99-130\\
HUTP-A061  \hfill hep-ph/9911406 \\
\vskip .3in
{\Large \bf Cosmology of Brane Models with Radion Stabilization}

\vskip 0.15in
{\bf Csaba Cs\'aki$^{a,}$\footnote{J. Robert Oppenheimer fellow.}, 
Michael Graesser$^b$, Lisa Randall$^{c,d}$\footnote{On leave of
absence from the Center for Theoretical Physics, Department 
of Physics, Massachusetts Institute of Technology, Cambridge, MA 02139.}  
and  John Terning$^e$}

\vskip 0.15in

$^a${\em Theory Division T-8, Los Alamos National Laboratory \\
Los Alamos, NM 87545}

\vskip 0.1in

$^b${\em Department of Physics\\
University of California, Santa Cruz, CA 95064}

\vskip 0.1in

$^c${\em Department of Physics\\
Princeton University, Princeton, NJ 08544}

\vskip 0.1in

$^d${\em Institute for Theoretical Physics\\
University of California, Santa Barbara, CA 93106}

\vskip 0.1in

$^e${\em Department of Physics\\
Harvard University, Cambridge, MA 02138}

\vskip 0.1in
{\tt  csaki@lanl.gov, graesser@scipp.ucsc.edu, \\
randall@feynman.princeton.edu, terning@schwinger.harvard.edu}

\end{center}

\vskip .15in
\begin{abstract} 
We analyze the cosmology of the Randall-Sundrum model and 
that of compact brane models in general in the presence
of a radius stabilization mechanism. We find that the expansion
of our universe is 
generically in agreement with the expected effective four 
dimensional description. The  
constraint (which is responsible
for the appearance of non-conventional cosmologies in these models)
that must be imposed on the matter densities on the 
two branes in the theory without a stabilized radius 
is a consequence of requiring 
a static solution even in the absence of stabilization. Such
constraints disappear in the presence of a stablizing potential, and the 
ordinary FRW (Friedmann-Robertson-Walker) equations are reproduced,
with the expansion driven by the sum of the physical values of the 
energy densities
on the two branes and in the 
bulk. For the case of the Randall-Sundrum model we examine the
kinematics of the radion field, and find that corrections to the
standard FRW equations are small for temperatures below the weak scale. 
We find that the radion field has renormalizable and 
unsuppressed couplings to Standard 
Model particles after 
electroweak symmetry breaking. These couplings 
may have important implications for collider searches. We comment on the
possibility that matter off the TeV brane could serve as a dark 
matter
candidate.

\end{abstract}
\end{titlepage}

\newpage
\section{Introduction}
\setcounter{equation}{0}
\setcounter{footnote}{0}
It has been widely understood over the past two years that a very promising
route towards reconciling the apparent mismatch of the fundamental scales 
of particle physics and gravity is to change the short 
distance behavior of gravity at 
distances much larger than the Planck length 
~\cite{Nima,otherextra,HW}. 
One prominent suggestion has been to lower the fundamental 
scale of gravity all the way to the weak scale by introducing large extra 
dimensions~\cite{Nima}. 
This possibility attracted much attention; in particular, the
presence of large extra dimensions opens up the possibility for new
cosmological scenarios for the early Universe, 
which has been the subject of study of several 
interesting 
papers~[4-11].
While the possibility of the existence of large extra 
dimensions is very exciting on its own, the existence of 
a big hierarchy between the weak and the Planck scales requires
the radius of the extra dimension 
to be much larger than its natural value (see 
\cite{CohenKaplan} for a scenario in which this might naturally occur).

Randall and Sundrum (RS) suggested a different 
setup~\cite{randall,randall2}, 
in which the extra dimensions are small, but the background metric is not
flat along the extra coordinate; rather it is a slice of anti-de Sitter
(AdS$_5$) space due to a negative bulk cosmological constant balanced by
the tensions on the two branes of this scenario. In this case, the curved 
nature
of the space-time causes the physical scales on the two branes to be 
different, and exponentially suppressed on the negative tension brane.
This exponential suppression can then naturally explain
why the physical scales observed are so much smaller than the Planck scale.
For early papers related to this subject see~\cite{noncompact,Gogberashvili};
generalizations of the RS models can be found 
in~[17-26],
embeddings into supergravity or string theory are discussed in [27-32].
Some of the 
aspects of the cosmology of these models have been examined 
in~[33-39],
and phenomenological consequences in~\cite{DDG,DHR}.

Clearly, the cosmology of this model can be very different from ordinary
inflationary cosmology in four dimensions. 
A more detailed study of
the early cosmology of the Randall-Sundrum model and that of 
brane models in general 
(including the large extra dimension scenario of \cite{Nima})
has been however hindered by one 
obstacle: it seemed that the {\it late} cosmology of brane models 
will generically deviate from the usual FRW Universe in the 4D theory on
our brane. This would bring into 
question most of the results of the early papers
about the cosmology of brane models, where it has been commonly assumed 
that the late cosmology (from BBN to present) of these models
is given by the standard FRW cosmology.
This conclusion is reached by 
applying the results of \cite{BDL}, who examined the solutions to
Einstein's equations in five dimensions on an $S^1/Z_2$ orbifold, 
with matter 
included on the two branes, and with no cosmological constants in the bulk
or on the brane. The result of this study was that the expansion of the 
brane does not reproduce the conventional FRW equations.
This result has been applied to the RS model in
\cite{CGKT,CGS}, and it was found that the cosmology of the model with
an infinitely large extra dimension (where one lives on the positive tension
brane, the ``Planck brane") in fact reproduces the ordinary FRW Universe. 
However, for the
model which solves the hierarchy problem (in which our universe lives on the 
negative tension brane, the ``TeV brane") there is a crucial sign 
difference in the Friedmann
equation. However, this conclusion about the wrong-signed Friedmann
equation (and the general conclusion that brane models have unconventional
late cosmologies) has  been reached in a theory without an explicit 
mechanism for
stabilizing the radius of the extra dimension, the radion field. It was 
already pointed out in \cite{GW,BDL,CGKT} that the effect of a radion 
stabilizing potential could significantly alter this picture. 

In this paper we confirm this intuition. We examine the cosmology of brane
models, in particular that of the 
RS model after the radius is stabilized (for example by the mechanism
suggested by Goldberger and Wise \cite{GW} or Luty and 
Sundrum \cite{LutySundrum}, 
but the details of the stabilization mechanism will be irrelevant).
We explain that the origin of the unconventional cosmologies is {\it not}
due to a breakdown of the effective 4D theory (neither is it due to the 
appearance of a negative tension brane in the case of the RS model), but
rather to a constraint that the matter on the hidden and visible branes 
are forced to obey in order to ensure a static radion 
modulus. This constraint ensures
a static solution for the radius even in the absence of a stabilization
mechanism. Once such a constraint on the matter content appears in a 
theory, all the results about non-conventional cosmologies mentioned 
above follow. However, if the radius is stabilized, such constraints disappear 
(they turn into an equation determining how the radius is 
shifted from its minimum due to the presence of matter on the branes),
and the ordinary FRW equations are recovered for the effective 4D theory
for temperatures below the weak scale. 
We also examine the kinematics and dynamics
of the radion field of the RS model in the presence of a 
stabilization force. 
We find that its natural mass scale is of the order of the weak scale, 
and that 
the shift in the radius is tiny for temperatures below the TeV scale. 
We also find that the radion TeV suppressed interactions with the 
Standard Model (SM), which results in renormalizable couplings to 
Standard Model fields after electroweak symmetry breaking. 
The existence of these couplings ensures that
the radion can decay before nucleosynthesis, and may also have 
important consequences for collider searches. 

We also comment on the possibility of matter in the bulk or on the Planck 
brane
serving as dark matter candidates. Matter in the bulk naturally has 
TeV suppressed couplings to SM fields, so its annihilation cross section
may result in interesting relic densities today. Matter on the 
Planck brane was presumably never in thermal equilibrium after inflation,
and it is therefore 
not clear whether it could have interesting relic densities.

This paper is organized as follows: 
in Section 2, we present the effective 4D description of the RS model.
We
explain the physical reasons why the constraint (which we derive 
in Appendix A together with an approximate solution in the bulk for
the case without a stabilized radion) should become irrelevant 
in the presence of a stable radion and find the effective 4D action for 
the radion-graviton system. In Section 3, we present a detailed 
solution to the 5 dimensional 
Einstein equations confirming the physical intuition of Section 2:
once the radius is stabilized, the ordinary FRW Universe is reproduced.
A similar solution to the case with vanishing background cosmological
constants is derived in Appendix A, illustrating that our arguments are
generic for brane models. In
Section 4, we discuss the kinematics and dynamics of the radion field 
based on the effective 4 dimensional action obtained in Section 2.
We also find the couplings of the radion field to the Standard Model
particles. 
In Section 5,  we discuss the issues related to (dark) matter on the hidden 
brane or in the bulk. 
We conclude in Section 6. Appendix A gives an approximate solution
to the Einstein equations for the RS model with matter on the branes, 
but no stabilization mechanism for the radion. Appendix B discusses 
the cosmology of brane models with vanishing  cosmological constants.
In Appendix C, we derive the radion mass in the Goldberger-Wise 
model for radius stabilization. In Appendix D, we show that the
Hubble law in the effective theory is consistent with the
Newtonian force law between particles on the TeV and Planck branes,
while in Appendix E we calculate the interaction strength of matter in the
bulk to matter on the Planck and the TeV branes. 

\section{The Effective Four Dimensional Theory}
\setcounter{equation}{0}
\setcounter{footnote}{0}
\label{section3}
In this section we construct the effective 4 dimensional equations of
motion of the RS model. We demonstrate, that the constraint on the
matter on the two branes (which is derived explicitly in Appendix A)
in the theory without a stabilized radius 
can also be obtained from the effective theory.
This constraint is a consequence of requiring a static solution to the 
equations of motion without a stabilization
mechanism. 

Throughout this paper (except in Appendix B) we consider the
RS model perturbed by matter on the two branes. The metric is given by
\begin{eqnarray} 
ds^2&=&n(y,t)^2 dt^2-a(y,t)^2 (dx_1^2+dx_2^2+dx_3^2)-b(y,t)^2 dy^2,
\nonumber \\
&\equiv& \tilde{g}_{AB}(x,y) dx^A dx^B.
\end{eqnarray}
The two branes are located at $y=0$ and at $y=1/2$. 
The Einstein tensor for this metric is given by 
\begin{eqnarray}
&& G_{00}=3\left[ \left(\frac{\dot{a}}{a}\right)^2
+\frac{\dot{a}\dot{b}}{ab}
-\frac{n^2}{b^2} \left(
\frac{a''}{a}
+\left(\frac{a'}{a}\right)^2 -\frac{a'b'}{ab}\right) \right], \nonumber \\
&& G_{ii}=\frac{a^2}{b^2} \left[ \left(\frac{a'}{a}\right)^2+
2 \frac{a'}{a}\frac{n'}{n} -\frac{b'n'}{bn}-2 \frac{b'a'}{ba}
+2\frac{a''}{a}+\frac{n''}{n}\right]+ 
\frac{a^2}{n^2} \left[-\left(\frac{\dot{a}}{a}\right)^2+2
\frac{\dot{a}}{a}\frac{\dot{n}}{n}-2\frac{\ddot{a}}{a}+\right.
\nonumber \\ 
&&
\left.\frac{\dot{b}}{b} \left( -2\frac{\dot{a}}{a}+\frac{\dot{n}}{n}
\right) -\frac{\ddot{b}}{b} \right],
\nonumber \\
&& G_{05}=3\left[ \frac{n'}{n}\frac{\dot{a}}{a}
+\frac{a'\dot{b}}{ab}-\frac{\dot{a}'}{a}\right],
\nonumber \\
&& G_{55}=3 \left[ \frac{a'}{a}\left(\frac{a'}{a}+\frac{n'}{n}\right)
-\frac{b^2}{n^2}
\left( 
\frac{\dot{a}}{a}\left(\frac{\dot{a}}{a}-\frac{\dot{n}}{n}\right)
+\frac{\ddot{a}}{a}\right)
\right].
\label{Einten}
\end{eqnarray}
Here primes (dots) denote derivatives with respect to $y$ ($t$). 
The Einstein equation is given by
$G_{ab}=\kappa^2 T_{ab}$, where $T_{ab}$ is the energy-momentum tensor,
and $\kappa^2=\frac{1}{2M^3}$, where $M$ is the five dimensional 
Planck scale. There
is a contribution to $T$ from the bulk cosmological constant of the form
\beq
T_{ab}^{bulk}=\tilde{g}_{ab} \Lambda,
\eeq
and from the branes 
\begin{eqnarray}
T_{a}^{b, brane}=
&&\frac{\delta (y)}{b} {\rm diag}\ (V_*+\rho_* ,V_*-p_*,V_*-p_*,V_*-p_*,0)
+\nonumber \\
&&\frac{\delta (y-\frac{1}{2})}
{b} {\rm diag} \ (-V+\rho ,-V-p,-V-p,-V-p,0),
\end{eqnarray}
where $V_*$ is the (positive) tension of the ``Planck'' 
brane at $y=0$, $\rho_*$ 
and $p_*$ are
the density and pressure of the matter included on the positive tension 
brane 
(where we assume an equation of state $p_*=w_* \rho_*$) as 
measured with respect to the metric $\tilde{g}$, 
and $\rho$ and $p$ are the density and pressure of the 
matter on the negative tension brane (the ``TeV'' brane), 
again measured with respect to  $\tilde{g}$. Hereafter quantities 
measured with respect to the metric $\tilde{g}$ are refered 
to as ``bare '' quantities. 
We know that in the limit $\rho ,p,\rho_*,p_*\to 0$ we want to recover the
static Randall-Sundrum solution\footnote{Note that in this 
paper we use a different notation than Refs.~\cite{randall,randall2}.
The parameter $k$ of \cite{randall,randall2} is denoted here by 
$m_0$, while
the radius of the extra dimension $r_c$ is denoted by $b_0$. 
For the coordinate along the extra dimension we use $y$ instead of $\Phi$,
and for the 5D Planck scale we use $M^3$ instead of $2 M^3$. 
In our notation $\kappa^2=1/2 M^3$. 
We also place the branes at $y=0$ and at $y=\frac{1}{2}$ instead of 
$0$ and $\pi$.}
of the form
\beq 
n(y)=a(y)=e^{-|y|m_0 b_0},
\eeq
where the relations between $\Lambda ,V_*$, $V$  and $m_0$ are given by
\begin{eqnarray}
&& V_*=\frac{6m_0}{\kappa^2}= -V, \nonumber \\
&& \Lambda =-\frac{6m^2 _0}{\kappa^2}.
\label{twofinetunes}
\end{eqnarray}
The effective 4D Planck scale is then given by
\beq
(8 \pi G_N)^{-1}=
M_{Pl}^2 \equiv \frac{1-\Omega^2_0}{\kappa^2 m_0} \hbox{,}
\label{mpl}
\eeq
where the notation for the present-day value of the warp factor  
\beq
\Omega_0 \equiv e^{-m_0b_0/2}
\eeq
has been introduced. 

Investigation of the cosmology of brane models has shown that in order
to find solutions to Einstein's equations with matter on the branes
there appears to be a constraint between the matter on the two branes, if
one requires that the extra dimension remains static \cite{BDL,kaloper}. 
The appearance of such constraints, as explained in Appendix A for the case 
of the RS model, generically leads to non-conventional
cosmologies. 
The constraint between the ``bare'' matter density $\rho_*$ on the 
Planck brane, and ``bare'' matter density $\rho$ on the TeV brane 
for the case of the RS model is found in Appendix A to be\footnote{For the
case of one extra dimension with two branes at $y=0$ and $y=1/2$ 
and with vanishing cosmological constants
this constraint
is of the form $\rho_* a(0)=-\rho a(\frac{1}{2})$ \cite{BDL}.} 
\begin{equation}
\rho_* = - \rho \Omega^2 _0.
\label{ft10}
\end{equation}
In addition, the computation in Appendix A of the Hubble parameter,
$H$, for the induced metric at the Planck brane 
implies that $\rho_* >0$. So the above constraint implies that 
the energy density on the TeV brane must be negative. 
Since this is at odds with phenomenology, it is 
important to understand the physical origin of these constraints before
concluding that this model (and brane models in general)
is phenomenologically unacceptable. 

In Appendix A it is shown that although there is a 
constraint between the Planck brane and TeV--brane energy densities, an 
effective 4D theory does exist. For example, for 
small perturbations in the 
energy densities the two branes expand at practically the 
same rate, and by an amount that agrees with the effective theory 
expectation.
This is consistent with what one expects in a 4D effective theory 
where there is a (approximate) uniform expansion of the two branes. 
This suggests that the origin of the unconventional cosmologies 
is the above--mentioned constraint, rather than the fact that 
the TeV brane has negative tension, or that there is a sick 
4D effective theory. 

In this section the 4D effective theory is constructed from two equivalent 
approaches. First, we directly average the 5D Einstein equations over the 
bulk to obtain some 4D equations. Alternatively, we average the 5D Einstein 
action 
over the bulk to obtain a 4D effective action. From both approaches we 
obtain the following picture: 
without a radion potential and for
generic energy densities on the two branes, the radion runs off to infinity. 
This can be avoided, again without a radion potential, by tuning the two 
energy densities in precisely the manner prescribed by Eq.~(\ref{ft10}).
Thus the above constraint equation is just a 
consequence of requiring that the radion modulus remains static in the 
absence of a stabilizing potential. Once the radion is stabilized the 
above constraint becomes irrelevant, and one obtains the conventional
late cosmology.

\subsection{The Averaged Einstein Equations}

In this subsection we average the 5D Einstein equations 
over the bulk in order to demonstrate that without a stabilizing 
potential the system is over--constrained once we require 
that the radion modulus is static. 
To perform the averaging, we linearize the metric about the 
RS solution:
\begin{eqnarray}
a(y,t)&=&a(t) \Omega(y,b(t)) \left(1+ 
\delta \bar{a}(y,t) \right) \nonumber \\ 
n(y,t)&=&\Omega(y,b(t))\left(1+\delta \bar{n}(t,y) \right) 
\nonumber \\
b(t,y)&=&b(t) \left(1+ \delta b(y,t)\right) .
\label{perturb1}
\end{eqnarray}
Here we keep the time dependence $b(t)$ in the warp factor $\Omega$,
and we will use the notation 
\beq
\Omega \equiv \Omega(y,b(t)) = e^{-m_0 b(t) |y|/2} \hbox{  , } 
\Omega_b \equiv \Omega(1/2,b(t)) \hbox{ .} 
\eeq
The value of $\Omega_b$, when $b=b_0=constant$, is then given by $\Omega_0$. 
In this expansion we allow $\delta f_i=$
$\delta \bar{a}$, $\delta \bar{n}$, $\delta b$ to be of the form 
$\delta f=\delta 
f(\rho(t),\rho_*(t),y)$  
and we only assume that $\delta f\sim O(\rho, \rho_*)$. 
This last assumption is  
reasonable, for in the limit that $\rho$, $\rho_* \rightarrow 0$ we should 
recover the RS solution. 
Thus an expansion in $\delta f_i $ is equivalent to an expansion in 
$\rho$, $\rho_*$. 
It is then sufficient to work to linear order in $\delta f_i$.
Terms such as $d \delta a /dt \sim \dot{\rho} \sim \rho^{3/2}$ are higher 
order in $\rho$, $\rho_*$ and are neglected, 
while we work to all orders in $b(t)$ and its time derivatives.

We also include a radion potential 
\beq
a^3  V_r(b) \equiv - b(t) \int dy \Omega^4 {\cal L}_R=
b(t) \int dy \Omega^4 T^{0} _0
\eeq
in the computations presented here, and set $V_r=0$ when desired.
Here ${\cal L}_R$ is some non--specified bulk dynamics responsible 
for generating a potential. 
Although our focus in this section is the constraint obtained without 
a radion potential, we include it here for later application.

Recall that 
in classical electromagnetism on a manifold without boundary one integrates 
$\nabla \cdot E = \rho$ over the manifold to conclude that the sum of 
the charges must vanish. So here, to see if there is a 
{\em topological }constraint on the 
component energy densities, we might try integrating
the analog of Gauss' Law in Einstein's theory. Namely, consider 
\begin{equation}
 \int dy \hbox{ }\Omega^4 G^0 _0 = \kappa^2 
\int dy \hbox{ } \Omega^4  T^0 _0 \hbox{ . } 
\label{constraint}
\end{equation}
We shall see that this equation does not lead to a topological 
constraint. Rather, this equation, combined with the average of 
the other Einstein equations demonstrates that the constraint on 
the energy densities follows from requiring a static extra dimension 
even when there is no radion potential.

Substituting the above expansion, Eq.(\ref{perturb1}), 
into Eq.~(\ref{constraint}) and 
integrating gives 
\begin{equation}
\frac{\dot{a}^2}{a^2}+ (m_0 b) \frac{\Omega^2_b}{1-\Omega^2_b} 
\frac{\dot{a}}{a} \frac{\dot{b}} {b} - \frac{(m_0 b)^2}{4} 
\frac{\Omega^2 _b}{1- \Omega^2 _b} \frac{ \dot{b}^2}{b^2}= 
\frac{\kappa^2 m_0} {3}\frac{1}{1-\Omega^2_b}
\left(\rho_* +\rho \Omega^4 _b +V_r(b)\right) + 
\epsilon^2  
\label{hubble} 
\end{equation} 
where $\epsilon^2 = O(\left(\delta \bar{a} \right)^2,\left(\delta \bar{n} \right)^2, 
 \left(\delta b \right)^2)$. 
Note that there 
are no corrections to $H^2$ linear in the perturbations. Thus the 
corrections to the Hubble formula from the details of the bulk geometry 
are important only at high temperatures. Also note that 
this equation reduces to the conventional FRW Hubble law
 when the energy density in the
radion 
is small (i.e. $\dot{b}$ and $V_r$ are negligible) 
compared to the other sources. 
(Recall that  
$8 \pi G_N= \kappa^2 m_0 / (1-\Omega^2_0)$).
Further, in this limit the 
expansion rate is set 
by 
$\rho_*+ \rho \Omega^4 _b$, which is the 
sum of the physically measured energy densities on 
the Planck and TeV branes, respectively.  
For small oscillations of the radion, 
the expansion of the universe is ordinary FRW.
Thus the bulk average of the $G_{00}$ equation 
resulted in the 4D Hubble Law.
 
Repeating the above averaging procedure for  
the $G_{ij}$ equation gives, for $\dot{b}_0 =0$ : 
\begin{equation}
\frac{\dot{a}^2}{a^2}+ 2 \frac{\ddot{a}}{a} = 
-8 \pi G_N \left(p_* +p \Omega^4 _0 - V_r(b_0)\right) + \epsilon^2
\label{gij}
\end{equation}
This is just the FRW ''pressure '' equation. As with the 
$G_{00}$ equation, there are no corrections of $O(\epsilon)$. 

The {\em unaveraged} linearized   
$G_{55}$ equation is given by  (again for $\dot{b}_0 =0$) : 
\beq
3 \frac{ \Omega '}{ \Omega} (3 \delta a ' + \delta n ') 
-3 \frac{b^2_0}{\Omega^2}  
\left(\frac{\dot{a}^2}{a^2}+ \frac{\ddot{a}}{a} \right) =
- \kappa^2 b^2 \left( 2 \Lambda \delta b  - T^5 _5  \right) 
+ \epsilon^2 \hbox{ .}
\label{t55}
\eeq 
Here we { \em do not } average over the fifth coordinate. If this is 
performed, then on the LHS quantities such as $\delta a$ and $\delta n$ 
appear, 
whose value depends on the detailed form of the solution in the bulk. 
The results of Appendix A could then 
be used to obtain the same  
constraint obtained below in Eq. (\ref{resulthere}). 
As seen below, more generality is obtained by not
averaging the $G_{55}$ equation. 
Next the ``jump equations'' for $\delta a'$ and $\delta n'$, derived 
in~\cite{BDL}, given by
\beq
\delta a' |_0  = - \frac{1}{6} \kappa^2 b  \left( \rho_* + \delta b V_* \right)
\hbox{ ,   }
\delta a' |_{1/2}= \frac{1}{6} \kappa^2 b \left( \rho + \delta b V \right )
\hbox{ ,}
\eeq
\beq
\delta n' | _0 = \frac{1}{6} \kappa^2 b \left(3p_* + 2 \rho_* - \delta b V_* 
\right) 
\hbox{ ,   }
\delta n' |_{1/2} = - \frac{1}{6} \kappa^2 b \left(3 p + 2 \rho - \delta b V 
\right)
\hbox{ ,}
\eeq
are inserted into Eq. (\ref{t55}) to obtain (for example at $y=0$)
\beq
- \frac{1}{6} \kappa ^2 m_0 (-\rho_* + 3p _* ) 
-  \left(\frac{\dot{a}^2}{a^2}+ \frac{\ddot{a}}{a}\right)
= - \frac{1}{3} \kappa^2 T^5 _5 | _0  \hbox{ .}
\label{t552}
\eeq

 
In the following we show that without a radion potential the 
 three equations (\ref{hubble}), (\ref{gij}) and (\ref{t552})
imply a constraint between the matter densities 
if a static solution $(b=constant)$ is imposed.  
The important point about Eq.~(\ref{t552}) is that in the absence 
of a radion potential
the system is over--constrained. To see this, we can 
eliminate the $a(t)$ dependence by using the other two 
equations, Eqs. (\ref{hubble}) and (\ref{gij}). For a static 
solution for $b$, i.e. $\delta b=0$ and $\dot{b}_0=0$, and with 
$V_r=T_{55}=0$,
this gives
\beq
- \frac{1}{6} \kappa ^2 m_0 (-\rho_* + 3p _* ) =\frac{1}{6}
\frac{\kappa^2 m_0}{1 - \Omega^2 _0} \left( \rho_* -3 p_* + 
 (\rho - 3 p) \Omega^4 _0  \right)
\eeq 
which simplifies to 
\begin{equation}
(-3 p + \rho) \Omega^4 _0 = (3 p_* - \rho_*) \Omega^2 _0.
\label{resulthere}
\end{equation}
In order that this equation 
remains consistent with the two conservation 
of energy equations, Eqs. (\ref{energy1}) and (\ref{energy2}), 
a further fine tuning   (for $w \neq 1/3$) $w=w_*$
is needed. So the above constraint is then 
\beq
\rho_* =- \Omega^2 _0 \rho \hbox{ ,}
\label{c2}
\eeq
which is the same 
constraint obtained from the explicit solutions Appendix A, Eq.~(\ref{c1}). 
Alternatively, the same constraint could have been obtained by 
comparing the {\em unaveraged} $G_{55}$ equations at $y=0$ and 
$y=1/2$ when $T_{55}=0$. 

This discussion then demonstrates the origin of the constraint: 
it is a consequence of requiring $b=constant$ 
without a radion potential. With these assumptions the system is 
over-determined, and a fine tuning of the energy densities is required 
to maintain a static solution in the bulk.

\subsection{The Effective Four Dimensional Action}

Below we derive the effective action for the four dimensional theory.
We will demonstrate 
that the above explanation for the origin of the constraint equation 
can also 
be obtained from this 4D action. This should be expected, since 
without a stabilizing potential the radion modulus is massless and 
appears in the 4D effective theory. Consequently, the
physics of maintaining a static modulus should also be obtained in the 
effective theory.  


To obtain the effective action, in 
the following we perturb the metric about the RS solutions as in 
Eqs. (\ref{perturb1}). 
We compute to $O(\delta f)$ and to all order in $b(t)$. Terms such 
as, e.g.  $d \delta a /dt  \sim \dot{\rho} \sim \rho^{3/2}$, are higher 
order 
in $\rho$, $\rho_*$ and are neglected. The effective Lagrangian is given
by
\begin{eqnarray}
{\cal L}_{eff} &=& -M ^3 \int^{1/2} _{-1/2} dy \sqrt{g} (R+ \Lambda/M^3) 
\nonumber \\
& & -a ^3 V_r(b)+ a^3 {\cal L}_{Pl}(a, \Psi^{(Pl)}_0 ,...) 
+a^3 \Omega_b^4 {\cal L}_{TeV}(\Omega _b a, \Psi^{(TeV)}_0, ... )\ ,
\end{eqnarray}
where  $V_r(b)$ is the radion potential, and 
${\cal L}$ denote the Lagrangians of the matter fields on the Planck and 
TeV branes,
expressed in terms of the bare (not rescaled) fields, 
bare masses and induced metric.
With our approximations the 4D effective action is then  
computed to be 
 \begin{eqnarray}
S_{eff} &=& -\frac{3}{ \kappa^2 m_0} \int dt a^3 
\left((1-\Omega^2_b)\frac{ \dot{a} ^2}{a ^2} + m_0 b \Omega^2 _b 
\frac{\dot{a}}{a} \frac{\dot{b}}{b} - 
\frac{(m_0 b)^2}{4} \Omega^2 _b\frac{\dot{b}^2 }{b^2 }\right) 
 -\int dt a^3 V_r(b)  \nonumber \\
& & +\int dt a^3 {\cal L}_{Pl} 
+\int dt a^3 \Omega_b^4 {\cal L}_{TeV}.
\label{4deff}
\end{eqnarray}
We have expressed the induced metric on the two branes as 
$\tilde{g}_{\mu \nu}= \Omega^2 _b \, {\rm diag}(1,-a^2,-a^2,-a^2)$. 
Since the curvature scalar
\beq
{\cal R}_{(4)}(a)= -6\frac{\ddot{a}}{a}-6 \left(\frac{\dot{a}}{a}\right)^2
\eeq
integrating the second term by parts results in a 
more conventional looking action
\begin{eqnarray}
S_{eff} &=& 
- 
\frac{1}{2 \kappa^2 m_0} \int dt a^3 (1-\Omega^2_b) {\cal R}_{(4)}(a)
+ \int dt a^3 \left(
\frac{3}{4} \frac{1}{\kappa^2 m_0} 
 (m_0 b)^2 \Omega^2 _b \frac{\dot{b}^2}{b^2}
- V_r(b) \right)
\nonumber \\ 
& & +S^M _{Pl}+ S^M _{TeV} 
\label{coveff}
\end{eqnarray}
where $S^M$ are the matter actions on the two branes.

It is now straightforward to compute the $b$ equation of motion.  
The point is that due to the dependence of $\Omega_b$ on 
$b$, the presence of the matter on the branes generates a 
potential for $b$. To see this, we compute the variation of 
the above action with respect to $b$, noting that (we assume) 
$S^M$ 
depends on $b$ only through the warp factor $\Omega_b$. Thus 
the contribution of the matter fields to the $b$ equation is 
\beq
\frac{\delta S^M}{\delta b} = \frac{\delta S^M}
{\delta \tilde{g}^{\mu \nu} } \frac{ \delta  \tilde{g}^{\mu \nu}} 
{\delta b} 
= - \sqrt{\tilde{g}}  \tilde{T}_{\mu \nu} \tilde{g}  ^{\mu \nu} 
 \frac{\Omega_b '}{ \Omega _b}  
= - \sqrt{g}  \frac{\partial}{\partial b}\frac{1}{4} \tilde{T} \Omega_b ^4 
 \ ,
\eeq
where $\tilde{g}_{\mu \nu}= \Omega^2 _b \, {\rm diag}(1,-a^2,-a^2,-a^2)$,
and
$\tilde{T}$ is the trace of the stress tensor\footnote{Thus we see
that $b$ couples to the trace of the stress tensor, and has ``dilaton''-like
couplings.} in terms of the 
bare fields and bare masses, and is equal to $\rho -3 p$ for a perfect fluid.
Thus the matter fields generate an effective potential for $b$ that is 
\beq
V_{eff}(b)=\frac{1}{4} \left(\rho_*- 3p_* + (\rho -3 p) \Omega^4 _b \right)
\ ,
\eeq
where we have added for later convenience 
a $b$ independent contribution from the Planck brane 
(which does not contribute to the $b$ equation). 

For general $\rho$ and $\rho_*$, the minimum of 
this potential is at $b \rightarrow \infty $. In fact with 
$\dot{b} =0$ and $V_r=0$, the $b$ 
equation is 
\begin{equation}
-\frac{3 \Omega^2_b}{ \kappa^2} \left( \frac{\ddot{a}}{a} +2 
\left(\frac{\dot{a}}{a}\right)^2 \right)
=-\frac{m_0}{2}( 3p -\rho) \Omega^4_b - \frac{3 \Omega^2 _b}{ \kappa^2}  
\left(\frac{\dot{a}}{a}\right)^2 \hbox{ ,}
\end{equation} 
which simplifies to 
\beq 
(3p-\rho) \Omega^4_b = 
\frac{6 \Omega^2_b}{\kappa^2 m_0}\left( \frac{\ddot{a}}{a} 
+\left( \frac{\dot{a}}{a}\right)^2 \right) \hbox{ .} 
\eeq
We note that this is just the average of the 
$G_{55}$ equation at the 
location of the TeV brane. Using  Eqs.~(\ref{hubble}) and 
(\ref{gij}) the above equation
 simplifies to 
\beq
(3p-\rho) \Omega^4_b 
= - \frac{\Omega^2_b}{1-\Omega^2_b} \left( 3 p_* -\rho_* 
+(3p-\rho)\Omega^4_b \right) \hbox{ .} 
\end{eqnarray}
For generic $\rho$ and $\rho_*$ the solution to this equation is 
$\Omega_b \rightarrow 0$. Since $\Omega_b =e^{-m_0 b/2}$, 
$b \rightarrow \infty$.
That is, the branes want to blow apart. There is another solution, however, 
and that is to 
allow a fine tune between $\rho_* $ and $\rho$.  
In fact, inspecting the above equation implies 
that $3p_* -\rho_* =- \Omega^2_b 
(3p -\rho)$ is also a static solution. 
Combining this with the conservation of 
energy equations for $\rho$, $\rho_*$, 
then implies the earlier constraint $\rho_* = - \rho \Omega^2 _b$. 

So this demonstrates that the constraint $\rho_* =- \rho \Omega^2_b$ is 
quite general, as it does not depend on the details of the 
solutions in the bulk. More importantly, this effective action 
computation explicitly shows that the constraint
directly follows from requiring that the 
radion modulus is static even when there is no stabilizing potential. From 
this perspective it is clear that with a radion potential 
$\rho_*$ and $\rho$ will not be required to be correlated. For uncorrelated 
$\rho$ and $\rho_*$, the branes want to go off to infinity; this however, 
will be balanced by the restoring force from the radion potential. What 
was once a constraint equation for $\rho$ and $\rho_*$, in the 
presence of the radion potential becomes an equation determining the 
new equilibrium point: $\Delta b \sim O(\rho, \rho_*)$.

\section{The Solution to the Einstein Equations in the presence of a 
Stabilizing Potential}
\setcounter{equation}{0}
\setcounter{footnote}{0}
\label{section4}
We have seen in the previous section that the constraint on the 
matter on the two branes is a consequence of the fact that without
radius stabilization one has an extra light field in the spectrum, 
whose equations of motion are static only for correlated values
of the matter densities. 
In this section, we will show that the conclusions reached 
above are in agreement with the detailed solutions of the 5 dimensional
Einstein equations, once the radius is stabilized. To study this
problem, we assume that a five dimensional 
potential $U(b)$ has been generated in the
5D theory by some mechanism (for example see \cite{GW}). Then the 
equations of motion in the bulk are given by
\begin{eqnarray}
&&G_{00}= \kappa^2 n^2 \left( \Lambda +U(b)\right), \ \ 
G_{ii}=-\kappa^2 a^2 \left(\Lambda+U(b)\right), \nonumber \\
&&G_{05}=0, \ \ 
G_{55}=-\kappa^2 b^2 \left( \Lambda +U(b)+bU'(b)\right), 
\label{eq:withpotential}
\end{eqnarray}
with $G_{AB}$ given in \ref{Einten}.
To simplify the solution of the equations, we assume that the mass of the
radion is very heavy, and that near it's minimum
$U$ is given approximately by
$U(b)=M_b^5(\frac{b-b_0}{b_0})^2$, where $b_0$ is the stabilized 
value of the 
radius, and the parameter $M_b^5$ is proportional to the radion
mass $m_{radion}$.
In order to understand what is happening to these 
equations we assume for a moment that 
 $M_b$ is the largest mass scale in the theory. Then the solution to
the last equation in Eq.~(\ref{eq:withpotential}) is simply given by 
$b=b_0$, with no other constraint on $a$ and $n$. With this solution we
also find that $U(b_0)$=0. Thus in the presence of a heavy radion field
the relevant Einstein equations are the $00$ and $ii$ components, with
the radius fixed to be at the stable value $b_0$. This already shows how
the constraint is eliminated. One of the equations of motion, which played
a vital role in establishing the correlation between the matter on the two 
branes is simply not appearing, it is automatically satisfied in the 
presence of the stable radius. 

The physics of this is that if the 
radion is very heavy, it can always adjust in an infinitesimal way to
satisfy this equation. As an analogy one can think of two charged 
spheres which  
gravitationally attract but electrically repel each other.
There will be a static solution only, 
if the force from the electric charges exactly
cancels the force from the gravitational attraction -- thus there is a 
relation between the masses and charges of the two spheres. If we
connect a spring with an extremely large spring constant between the 
two spheres, however, the only 
condition we will have is that the spring be at its own equilibrium point,
and the other forces are canceled by infinitesimal changes in the length 
of the spring.

Thus what we are left to show is that the first two equations, with the 
radion set to $b_0$ have solutions for arbitrary values of matter 
perturbations. To simplify the calculation, we perturb around the 
RS solution with cosmological constants $\delta V$ and $\delta V_*$
instead of matter densities. The appropriate Hubble law for matter densities 
can be obtained by the substitution $\delta V\to \rho$, 
$\delta V_*\to \rho_*$ and $H \to \dot{a}_0(t)/a_0(t)$ (to lowest order 
in the perturbations). 
We solve the equation using the ansatz
\beq a(t,y)=e^{Ht} e^{-|y|m_0b_0} (1+\delta a(y)), \ \ \
n(t,y)=e^{-|y|m_0b_0} (1+\delta a(y)), \ \ \ b=b_0.
\eeq
This way the $00$ and the $ii$ Einstein equations reduce to a single 
equation of the form 
\beq
H^2b_0^2= a(y) a''(y) +a'(y)^2-2m_0 ^2b_0 ^2 a(y)^2,
\eeq
where $a(y)=e^{-|y|m_0b_0 } \left( 1+\delta a(y)\right)$. 
Plugging back this ansatz and 
expanding to first order in $\delta a(y)$ we obtain the equation
\beq
\delta a''(y)-4 b_0 m_0 \delta a'(y)= b_0 ^2 H^2 e^{2|y|m_0b_0}.
\label{perteq}
\eeq
This equation is a linear inhomogeneous ordinary differential equation,
which can be solved by the standard rules. The general solution
is given by
\beq
\delta a(y)= \frac{\alpha}{4m_0 b_0} 
(e^{4 |y|b_0 m_0}-1)-\frac{H^2}{4 m_0 ^2} (e^{2 |y|b_0m_0}
-1),
\label{da}
\eeq
where the overall constant has been fixed such that $\delta a(0)=0$.
The remaining two constants, $\alpha$ and $H^2$ have to be fixed such 
that the jumps of this function at the two branes reproduce the 
matter perturbation that we are including. The result is given by
\beq
H^2=\frac{\kappa^2 m_0}{3 (1-\Omega^2_0)} (\delta V_* +\delta V \Omega^4_0),
\label{hh}
\eeq
and the value of the other constant $\alpha$ is given by
\beq
\alpha =\frac{\kappa^2 b_0}{6 (1-\Omega^2_0)} (\delta V_* \Omega^2_0
+\delta V  \Omega^4_0).
\label{alpha}
\eeq
We note that Eq.~(\ref{hh}) is the standard Hubble law formula with correct 
normalization for the 
physically observed energy density $\delta V_* + \Omega^4_0 \delta V$.
The consistency of this solution requires $\delta a\ll1$, which by 
inspection 
implies that $\delta V_*$ is much less than the intermediate 
scale $\Omega^2 _0 M^4 _{Pl}$, 
and that $\delta V \Omega^4_0 \ll ({\rm TeV})^4$. Thus the 4D effective
theory should be valid when these conditions are satisfied, in particular
if the temperature on the TeV brane is below the TeV scale.

The lesson is as expected: radion stabilization removes the 
constraint between the matter fields on the two branes, there is a solution
to Einstein's equations for any value of the perturbations, and the 
Hubble constant is given by the expression expected from the effective 4D 
description of the theory. To show that our conclusions are generic for
brane models, we derive a solution similar to the one presented in this 
section
in Appendix A for the case of vanishing cosmological constants 
(the case considered in \cite{BDL}) in the 
presence of a stabilized radius.

\section{Radion Kinematics and Dynamics}
\setcounter{equation}{0}
\setcounter{footnote}{0}
\label{section5}

In this Section we concentrate on the radion field and point out 
some implications of the 4D effective action, Eq.~(\ref{4deff}).
First we discuss the cosmological consequences of (\ref{4deff}),
and then we examine the couplings of the radion to the SM fields and
the consequences of these couplings.

\subsection{Radion Cosmology}
Recall that for  
the metric (dropping the $\delta a$, $\delta n$ and 
$\delta b$ perturbations of Section \ref{section3}, since they only 
contribute at $O((\delta f)^2) $ )
\begin{eqnarray} 
a(t,y) &=& a(t) e^{-m_0 b(t) |y|}  \nonumber \\
n(t,y) &=& e^{-m_0 b(t) |y|} \nonumber \\
b(t,y) &=& b(t)  
\end{eqnarray}
the 4D effective action is 
\begin{eqnarray}
S_{eff}&=& -\frac{3}{ \kappa^2 m_0} \int dt a^3 
\left((1-\Omega^2_b)\frac{ \dot{a} ^2}{a ^2} + m_0 b \Omega^2_b 
\frac{\dot{a}}{a} \frac{\dot{b}}{b} - 
\frac{(m_0 b)^2}{4} \Omega^2 _b\frac{\dot{b}^2 }{b^2 }\right) 
 -\int dt a^3 V_r(b)  \nonumber \\
& &  +S^M _{Pl}+ S^M _{TeV} \ .
\end{eqnarray}
  From this equation we learned in Section 2.2
that without a stabilizing potential a 
constraint must be imposed on the different energy densities to maintain 
a constant radion modulus. In this Section we include a stabilizing 
potential 
$V_r$. 

The presence of the interaction $\dot{a} \dot{b}$ implies that in this 
coordinate basis the  
physical radion is mixed with the massless graviton. To separate the 
fields, it is convenient to perform a conformal transformation on the 
metric:
\begin{eqnarray}
a(t) &=& f(b(\bar{t})) \hbox{ } \bar{a} (\bar{t}) \nonumber \\
d t &=& f(b(\bar{t}))  \hbox{ }d \bar{t}  \hbox{ . }
\label{wrescale}
\end{eqnarray}
We want the action in the new basis 
to contain no cross terms $\dot{a} \dot{b}$. This fixes $f$ to be 
\beq
f(b) = \left( \frac{1- \Omega^2 _{0}}{1-\Omega^2_b} \right)^{1/2} \hbox{ ,}
\eeq
where recall that $\Omega _{0}$ is the present--day value of $\Omega_b$.
Note that for small perturbations of $\Omega_b$ away from its present--day 
value, $f(b) =1- \Omega^2_0 
m_0 \delta b/2 + \cdots$. So for small perturbations, 
i.e. $m_0 \delta b \ll 1$, 
the difference between the 
Einstein frame and the original frame is incredibly small, 
$O(10^{-30})$.
The difference between the two frames is then only important for 
large departures in $\Omega_b$ from its current value.   

Then the 4D effective action in this frame is :
\begin{eqnarray}
S&=& \int  d \bar{t} \bar{a}^3 (\bar{t}) \left( 
-\frac{3}{\kappa^2 m_0}(1-\Omega^2_0)
\left(\frac{\dot{\bar{a}}^2}{\bar{a}^2} 
- \frac{1}{4} (m_0 b)^2 \frac{\Omega^2_b}{(1- \Omega^2_b)^2} 
\frac{\dot{b}^2}{b^2}
\right)
-f(b)^4 V_r(b) \right) \nonumber \\ 
& &  +\int d \bar{t} \bar{a} ^3 f(b)^4 
{\cal L}_{Pl}(\bar{a}, \Psi^{(Pl)} _0,...) +\int d \bar{t} \bar{a}^3 
f(b)^4 \Omega_b^4 {\cal L}_{TeV}( \Omega_b \bar{a}, \Psi^{(TeV)} _0,...).
\label{ef}
\end{eqnarray}
In this action the radion appears as a scalar field 
with a non--trivial ($b$--dependent) kinetic term. In the 
absence of sources the radion potential is 
\beq
\overline{V}_{r}(b)=  f(b)^4 V_r(b). 
\eeq
In the presence of sources one can show similarly to the arguments
in Section 2.2 that this is modified to 
\beq
\overline{V}_{r,eff} (b)=\overline{V}_{r}
 +\frac{f(b)^4}{4}\left( \rho_* -3 p_* 
+(\rho -3 p) \Omega^4_b \right) \hbox{ .}
\eeq 
It is also convenient to define a new scale 
\beq
\Lambda _W = \Omega_0 M_{Pl} \hbox{ .}
\eeq
Then $\Lambda_W \sim  {\cal O}($TeV$)$. Inspecting the above Einstein 
frame Lagrangian, we see that for small perturbations away from 
$\Omega_b=\Omega_0$, the canonically normalized radion $\phi$ is 
\beq
m_0 b(t) = \sqrt{\frac{2}{3}} \frac{\phi(t)}{\Lambda_W} (1-\Omega^2_0) 
\sim \sqrt{\frac{2}{3}} \frac{\phi(t)}{\Lambda_W} \hbox{ .}
\label{normb}
\eeq
In the 
following the $\Omega^2_0$ corrections are neglected, so the second 
expression on the RHS is used. 

In this frame the Hubble law has a simple expression and interpretation:
(This can also be obtained from performing the conformal 
rescaling in Eq.~(\ref{hubble}))
\begin{eqnarray}
\frac{\dot{\bar{a}}^2}{\bar{a}^2}&=& \frac{8 \pi G_N}{3} \left( 
f(b)^4( \rho_* + \rho_{vis}) +  \frac{1}{4} \frac{3}{8 \pi G_N} 
(m_0 b)^2 \left( \frac{\Omega_b}{1-\Omega^2_b}\right)^2 \frac{\dot{b}^2}{b^2}
+ \overline{V}_r(b) \right) \nonumber \\
&=&  \frac{8 \pi G_N}{3} \left( \frac{1}{2} \dot{\phi}^2+ 
\overline{V}_r(\phi) + f(b)^4( \rho_* + \rho_{vis})
\right) \hbox{ .}
\end{eqnarray}
Here and throughout 
$G_N$ is the present--day value of Newton's constant, and the 
second expression is only valid for $b$ close to  
$b_0$. 
We see from this equation that as 
long as the energy density in the radion does not dominate the 
energy density in $\rho_{vis} =\rho \Omega^4_b$, in the Einstein frame 
the 
universe expands as in the usual cosmology. 
Since by assumption the oscillations are small, in the original 
frame we also have 
FRW expansion since then $a(t) =\bar{a}(\bar{t})\times 
(1+ \hbox{small osc.})$. 

In the absence of sources $(\rho=\rho_*=0)$ we require that the 
potential truly stabilizes the radion. 
That is, we require 
$\dot{b}=0$ is a solution. Inspecting the $b$ equation we see that
a static solution in this case 
is possible only if $\overline{V}'=0$. 
If we also require that $a$ is static in this limit 
then the $a$ equation of motion implies that $\overline{V}=0$. 
Since $f \neq 0$ and $f' \neq 0$, at the local 
minimum we must have $V=0$ and $V'=0$. 
With this knowledge, the radion mass obtained from the above action is then
given by
\beq
m^2_r = \frac{2}{3 m^2_0 }\frac{\overline{V}_r''(b_0)}{M^2_{Pl}\Omega^2_{0}} 
(1-\Omega^2_0)^2
\hbox{ .}
\label{radionmass}
\eeq
Near the local minimum it is then convenient to expand $V_r$ as 
\begin{eqnarray}
V_r(b)&=& \frac{3}{4} m^2_r \left(\frac{m_0 b_0}{1-\Omega^2_0}\right)^2 
\Omega^2_0 M^2_{Pl} \left(\frac{b-b_0}{b_0}\right)^2 
\nonumber \\
&=& \frac{1}{2} m^2_r (\phi-\phi_0)^2 \hbox{ .}
\label{vr}
\end{eqnarray}
For the Goldberger--Wise stabilizing 
mechanism, $V'' \sim M^2_{Pl} \times O($TeV$)^4$ 
(see Appendix B), so 
from the above formula $m_r \sim O($TeV$)$. This is not surprising, since 
a bulk scalar field with $O(M_{Pl})$ bare mass appears in the 4D effective 
theory as a Kaluza--Klein tower of scalars, where the lightest field 
has an $O($TeV$)$ mass. 

In the following the action for the canonically normalized radion 
$\phi$ is then truncated at 
\beq
S_{radion}=  \frac{1}{2} 
\int d^4 x \sqrt{g} \left((\partial \phi)^2
 -m^2_r  ( \phi -\phi _0)^2 \right)  \hbox{ .} 
\eeq
This will be the appropriate action for describing small 
displacements of the radion from its minimum.

In Section \ref{section3} it was argued that the special constraint 
between the energy densities on the two branes disappears once the 
radion is stabilized. Using the Einstein frame action, Eq.(\ref{ef}), 
the $b$ equation is 
\beq
\frac{3}{16 \pi G_N} m^2_0 b \frac{\Omega^2_b}{(1-\Omega^2_b)^2} 
\left(\frac{\ddot{b}}{b}+3 \frac{\dot{\bar{a}}}{\bar{a}}\frac{\dot{b}}{b}
- \frac{m_0 b}{2} \frac{\dot{b}^2}{b^2} 
\frac{1+ \Omega^2 _b }{1-\Omega^2 _b} 
\right)  =
 -\overline{V}'_{r,eff}  
\label{bequation}
\eeq
This 
equation can be used to recover 
the earlier constraint $\rho_* =- \Omega^2_b \rho$ 
when $\overline{V}=0$ and $\dot{b}=0$ is 
imposed. Once this constraint is not satisfied, it is clear from inspecting 
the above equation that the restoring force term $\overline{V}'$ will 
balance the expansion force term. More concretely, plugging 
$V_r$,  Eq.~(\ref{vr}), into the above equation and neglecting the 
$\dot{b}$ and $\ddot{b}$ terms gives 
\beq
\frac{\Delta b}{b_0}= \frac{1}{3}
\frac{1-\Omega^2_0}{m_0 b_0} 
\frac{(\rho_{vis}-3p_{vis})}{m^2 _r \Omega^2_0 M^2_{Pl}} 
\label{db}
\eeq
where $\Delta b$ is the distance between the minimum of the effective 
potential with and without matter, and we assume that the radion 
was at its minimum at the start of matter--domination (MD). Also 
$\rho_{vis} =\Omega^4_b \rho$, etc. is the physically measured 
energy density on the TeV brane. 
To obtain this formula it was also assumed 
that the Planck brane is empty $(\rho_* =0)$. 
To include $\rho_* \neq 0$, in the above formula replace 
$\rho_{vis} \rightarrow \rho_{vis} + \Omega^2 \rho_*$, etc. 
The consistency of neglecting the $\dot{b}$ and 
$\ddot{b}$ terms requires $\Delta b/ b \ll 1$, for then 
$\dot{b} \sim \dot{\rho} \sim 
\rho^{3/2}$ is suppressed compared to the $\Delta b$ term. 

From Eq.~(\ref{db}) it is seen that 
there is no shift in $b$ 
during a radiation dominated era $(w=1/3)$ 
from the dominant component.\footnote{This is because the stress
tensor of radiation is traceless. In what follows we neglect the 
quantum corrections to Tr T.} Hence the physical implications
of this shift are much smaller than what one would have naively thought. 
During this era, however, the energy density contained a small component 
of non--relativistic matter, which will cause $b$ to shift. 
Since both Newton's constant and 
SM particle masses depend on the vev of the radion, a substantial 
shift in these quantities during any era after BBN would lead 
to a BBN somewhat different from the standard BBN cosmology. 
The success of standard BBN then implies that these changes 
must be small, and this is found to be the case here.   
To estimate the shift since the start of BBN, we note that
the energy density in non--relativistic 
matter just before the start of BBN was
$\rho _{NR} \sim (T_{BBN}/ T_0)^3  \rho_{c,0} \sim 10^{20} $ eV$^4$, 
where we have used $T_{BBN} \sim 10$ MeV, $\rho_{c,0} \sim 10^{-10}$ 
eV$^4$.
Substituting $\rho_{vis}- 3 p_{vis} \sim \rho_{NR}$ into   
Eq.~(\ref{db}) gives 
\beq
\frac{\delta G_N}{G_N} \sim \Omega^2_0 m_0 \delta b \sim  10 ^{-30 } 
\times 10^{-28} \left( \frac{\hbox{TeV}}{m_r} \right)^2  \hbox{ .} 
\eeq
We also require that the weak scale has not shifted by more 
than ${\cal O}(10 \%)$ since BBN, 
which implies $\Delta b / b < {\cal O} (1/ m_0 b_0)$ .
From Eq. (\ref{db}) this implies  
\beq
m_r > 10^{10}
\frac{(\hbox{eV})^2}{\hbox{ TeV}} \sim {\cal O}(10^{-2}) \hbox{ eV} .
\eeq
Of course, it is obvious from the discussion in Section 4.2 that 
accelerator experiments provide a much stronger constraint on 
the radion mass than the above result. 

One general drawback of the RS model seems to be that there are two 
fine-tunings required to obtain a static solution (see 
Eq. (\ref{twofinetunes})). One of these tunings
is clearly equivalent to the usual cosmological constant problem,
however, it seems that there is a second equally bad tuning appearing,
making the model less attractive. We will argue that 
(as already emphasized by \cite{GW,Raman})
the second
tuning was only a consequence of looking for a static solution without
a stabilized radion, and disappears in the 
presence of a stabilization mechanism. To see this, assume that the 
radius is stabilized, and we perturb the brane tensions as 
$V \rightarrow V + \delta V$ and $V_* \rightarrow V_* + \delta V_*$ so 
that the fine-tune relations Eq. (\ref{twofinetunes}) no longer hold. 
From the previous discussions about stabilizing the radion, 
we find that the first perturbation 
generates an effective potential for $b$: $\delta V_{eff} 
\sim \delta V \Omega^4 _b$. 
This perturbation has two effects: to shift $b$, and to increase the 4D 
cosmological constant. 
The 4D cosmological constant is canceled by appropriately 
choosing $\delta V_*$, so one fine--tuning equivalent to the 
4D cosmological constant problem remains. But if 
$\delta V$ is small compared to both $V$ and the typical 
mass scale in the stabilizing potential, then from Eq. (\ref{db}) we find 
that the shift in $b$ is tiny. Thus once the radion is stabilized, 
only one fine--tuning of ${\cal O}(M^{-4} _{Pl})$  is required 
(equivalent to canceling the 4D cosmological constant).
The other tuning required in order to get the correct hierarchy of 
scales is clearly only of  ${\cal O}(10^{-1}-10^{-2})$.  The 
conclusion is that there is
only one severe fine--tuning in the RS model, equivalent to the cosmological
constant problem. 

We next argue that there is no moduli 
problem associated with the radion.
A potential moduli problem can occur from two sources. 
The first is that coherent oscillations of the radion can 
overclose the universe well--before BBN, ruining the success 
of BBN. In the next section  
we will see that after EWSB, 
the radion has renormalizable interactions with 
the SM fields. So after EWSB it promptly decays, long before the 
start of BBN. So by the start of BBN there is no 
energy in the coherent oscillations of the radion. 
The second concern is that the above shift in 
$b$ may introduce a correction to the Hubble Law which would 
modify the time evolution of our universe. This too could spoil 
the success of standard BBN cosmology. We shall find that
the correction is sufficiently small that this is not a concern.

The shift in $b$ will add energy to the radion potential. As the 
universe cools this energy will 
not red shift like matter, since it is not a 
coherent oscillation. Rather, as the universe cools $b$ will follow its 
instantaneous minimum since the cooling is adiabatic (for $ m_r \gg H)$. 
The energy stored in the radion potential then 
appears as a $\rho^2$ correction 
to the Hubble law. This could affect the time evolution of 
the universe.  
For a general equation of state, we find 
from the above expressions for $\Delta b$ 
and $V_r$ that
\beq 
\frac{V_r(\Delta b)}{\rho_{vis}} \sim  
\frac{(3 p_{vis} -\rho_{vis})^2}{m^2_r \Lambda^2 _W \rho_{vis} } 
\sim \frac{\rho_{NR}^2}{m^2_r \Lambda^2 _W \rho_{vis}} 
\hbox{.} 
\eeq
Using the value of $\rho_{NR} \sim 10^{20}$ eV$^4$ when 
$T \sim$ 10 MeV, then the LHS is $\ll 1$ if $m_r >10^{-6} \hbox{eV}$ 
which is weaker than the collider constraints. 
So there is no moduli problem from the shift in the radion. 

We conclude this section by arguing that a stabilizing potential is in fact
required in order that the SM particle masses have not changed 
significantly since the start of MD. (Of course 
direct accelerator searches require that the radion is heavier 
than ${\cal O}$(10--100) GeV.) 
The point is 
that without a stabilizing potential, the source terms on the TeV wall
generate a potential for $b$, which wants the walls to separate.
We can estimate the rate of expansion for $b$ 
by setting the stabilizing potential to zero in (\ref{bequation})
(which does not mean that the full $\overline{V}_{eff}$ vanishes).
Inspecting that equation one can find that the $\dot{a}\dot{b}$ term 
never dominates the expansion, thus the radion is not slowly rolling.
One finds that the $\ddot{b}$ and $\dot{b}^2$ terms in (\ref{bequation})
are comparable, and this results in an expansion rate
\beq
\frac{\dot{b}}{b}\sim H\frac{M_{Pl}}{\Lambda _W}
\eeq
for $b$ close to $b_0$. Thus the fractional change in the radius will
be significant (${\cal O}(1)$) over time scales much shorter than the 
Hubble time, which results in significant changes in 
the SM particle masses during MD, completely changing the
predictions for BBN. Thus the success of BBN cosmology requires that the
radion is stabilized before the onset of BBN.

\subsection{Couplings of the Radion to SM Particles}

The radion field interacts with SM particles, and the size of these 
interactions determines whether the radion can decay quickly enough 
to avoid overclosure by the start of BBN.
We shall see that the radion has renormalizable and
and $1/$TeV suppressed 
interactions with the SM fields. 

In the RS scenario, the radion has renormalizable interactions 
with SM fields after electroweak symmetry breaking (EWSB).
This is because the mass scales 
on the TeV brane depend on the radion modulus by  
\beq
<h>=v= v_0 e^{-\frac{1}{\sqrt{6}} \frac{\phi}{\Lambda_W}} \hbox{ .} 
\label{vev}
\eeq
But since this warping contains the radion modulus, 
this dependence introduces renormalizable couplings of the 
radion to matter once EWSB occurs. For example, in the Yukawa interaction 
we expand the Higgs vev as in Eq.(\ref{vev}) to obtain
\beq
\lambda_{ij}  h \overline{\psi}_i \psi_j  \rightarrow 
\lambda_{ij} v  \overline{\psi}_i \psi_j
- \lambda_{ij} \frac{1}{\sqrt{6}} \frac{v}{\Lambda_W } \phi
\overline{\psi}_i \psi_j  + \cdots \hbox{ .}
\eeq 
So if present, this results in renormalizable interactions between the 
radion and the SM particles, scaled only by 
$v/ \Lambda_W$ in addition to the usual factors of 
Yukawa couplings or gauge 
couplings. 
Since the radion mixes with both the neutral Higgs 
and with the massless graviton (though the latter is negligible 
compared to the radion-Higgs mixing), 
we must work in a basis where the radion does not 
mix with these fields. 

Next the action for the radion and Higgs interactions is derived, and 
the physical eigenstates and their couplings 
to SM fields are identified. 
We parameterize the metric on the TeV brane as 
\beq 
\tilde{g}_{\mu \nu}(x,y=1/2) = \Omega^2_b(x) g_{\mu \nu}(x) \hbox{ , } 
\eeq
\beq
 g_{\mu \nu}(x)= \eta_{\mu \nu} + h_{\mu \nu}(x) \hbox{ ,}
\eeq
\beq
\Omega_b(x) =e^{-m_0 b(x)/2} 
\eeq
and in the bulk as 
\begin{eqnarray}
\tilde{g}_{\mu, \nu}(x,y)&=& e^{-2 m_0 b(x) |y|} g_{\mu \nu}(x) 
\hbox{ ,} \nonumber \\
g_{55}(x,y) &=&- b^2(x) \hbox{ .}
\end{eqnarray}
This expression then includes 
the fluctuations in the radion and the zero--mode 
graviton. 
Inserting this into the 5D action and integrating over the 
extra dimension results in the following 4D effective action 
\begin{eqnarray}
S_{eff}&=& \int d ^4 x \sqrt{g} 
\left(- \frac{1}{2} \frac{(1-\Omega^2_b)}{\kappa^2 m_0}  
{\cal R}_{(4)}(g) +  
\frac{3 m_0}{4 \kappa^2}  \Omega^2_b 
(\partial b)^2 -V_r(b) \right) +S_{TeV}+S_{Pl} \hbox{ .}
\label{cov4eff}
\end{eqnarray}
From this action 
we find that for small 
fluctuations about $b=b_0$, the canonically normalized radion 
$\phi$ is 
\beq
m_0 b(x) = \sqrt{\frac{2}{3}}\frac{ \phi(x)}{\Lambda_W}
\eeq
which is the same as Eq.~(\ref{normb}) for small $\Omega_0$.
(The small difference between this and 
 Eq.~(\ref{normb}) is due to the latter normalization being defined 
in the  
Einstein frame.)
The SM action on the TeV brane is
\begin{eqnarray}
S_{TeV} &=& \int d^4 x 
\sqrt{\tilde{g}} \left( \tilde{g}^{\mu \nu} D_{\mu} H^{\dagger} D_{\nu} H 
- V(H) + i\bar{\psi}_i \tilde{e}^{a \mu} \gamma_a D_{\mu} \psi_i
 - \lambda_{ij} H \psi_i \psi_j +h.c.
\right) \nonumber \\
&=&\int d^4 x \sqrt{g} \Omega^4 _b 
\left( \Omega^{-2} _b g^{\mu \nu} D_{\mu} H^{\dagger} D_{\nu} H 
- V(H) + i \Omega^{-1} _b \bar{\psi}_i e^{a \mu} \gamma_a D_{\mu} \psi_i
 - \lambda_{ij} H \psi_i \psi_j +h.c. 
\right) \hbox{ ,}\nonumber \\
\label{SMl1}
\end{eqnarray} 
where for simplicity only the fermion and Higgs 
interactions are included.
It is straightforward to include other interactions, for example, the 
Yang-Mills kinetic terms.
We assume a Higgs potential of the form 
\beq
V(H) = \lambda ( H^{\dagger} H - \frac{1}{2} v^2 _0)^2 \hbox{ .} 
\eeq
To obtain 
canonical normalization of the kinetic terms in Eq.(\ref{SMl1}) 
we rescale   
$H \rightarrow \Omega^{-1} _b  H $ and 
$\psi \rightarrow \Omega^{-3/2} _b \psi $ . 
The effect of this is to renormalize $v_0 \rightarrow \Omega_b v$, and 
to introduce higher dimension operators between the radion and the 
SM fields. That is, the new SM action is 
\begin{eqnarray} 
S_{eff} &=& \int d^4 x
\sqrt{g}  
\left( g^{\mu \nu} D_{\mu} H^{\dagger} D_{\nu} H 
- V(H, \phi) + i \bar{\psi}_i e^{a \mu} \gamma_a D_{\mu} \psi_i
 - \lambda_{ij} H \psi_i \psi_j+h.c. 
\right) \nonumber \\
& & +S_{radion}+ \int d^4 x  \sqrt{g} \left(\frac{1}{\sqrt{6}} 
\frac{\partial \phi}{\Lambda_W}
\left( H^{\dagger} D H +h.c.\right)+ 
\left(\frac{\partial \phi}{\sqrt{6} \Lambda_W} \right)^2 H^{\dagger} H 
\right)
\nonumber \\
& &
+ \int d^4 x \sqrt{g} \left(
\frac{3}{2 \sqrt{6} \Lambda_W} 
i \bar{\psi}_i \gamma^{\mu} \psi \partial _{\mu}  \phi \right) +S_{Pl}
\hbox{ .}
\label{faction}
\end{eqnarray}  
Note that in the kinetic terms 
the radion has dimension 5 and dimension 6
interactions with fermions 
and bosons. 
The Higgs potential after the rescaling is  
\beq
V(H,\phi)= \lambda \left( H^{\dagger} H- \frac{1}{2} v^2_0 
e^{- \sqrt{\frac{2}{3}}\frac{\phi}{\Lambda_W}}\right)^2   \hbox{ .}
\eeq
Note that this confirms the intuition that the source of the 
radion coupling to SM particles is the conformal breaking sector. 
This potential generically results in mass mixing between the Higgs and 
the radion. Inspecting the final action, 
Eq.(\ref{faction}), we also 
see that after EWSB there is kinetic mixing between the
 radion and Higgs.  
So in order to verify the claim that after 
EWSB the radion has renormalizable 
interactions with the SM fields, we must identify the 
coupling of the SM fields to the physical eigenstates made 
out of $H^0$ and $\phi$, and in particular verify that the 
gauge eigenstate $H^0$ is not 
a mass eigenstate. 

Expanding $H^0=(v+ h+i H_I)/\sqrt{2}$, the  
$\phi$ and $H^0$ equations are 
\beq
0=m^2_r (\phi -\phi_0) +\lambda \gamma v_0 
e^{- \sqrt{\frac{2}{3}}\frac{\phi}{\Lambda_W}} (v^2 - v^2_0
e^{- \sqrt{\frac{2}{3}}\frac{\phi}{\Lambda_W}} )
\eeq
and 
\beq 
0= \lambda v (v^2 -v^2_0
e^{- \sqrt{\frac{2}{3}}\frac{\phi}{\Lambda_W}}) \hbox{ ,}
\eeq
where the scaling factor $\gamma$ is defined by 
\beq
\gamma \equiv \frac{v}{\sqrt{6} \Lambda_W}= \frac{1}{\sqrt{6}}
\left( \frac{v}{\Omega_0 M_{Pl}} \right)~.
\eeq
Note that $\gamma$ is fixed by the precise 
value of $\Omega_0$. 
So the vacuum that breaks EWS is 
\begin{eqnarray}
< \phi > &=& \phi_0  \nonumber \\ 
v^2 &=& v^2_0
e^{- \sqrt{\frac{2}{3}}\frac{\phi_0}{\Lambda_W}} \hbox{ .}
\end{eqnarray}
This is physically reasonable, since the radion and Higgs 
potential is the sum of two positive definite quantities, 
which has a local minimum when each term separately vanishes. 
This also demonstrates that at tree--level 
the fermion and gauge boson mass spectrum is the same here as
in the SM, since 
the vev of $\phi$ only determines $v$, and otherwise does not 
contribute to any of those
masses.

In the original gauge basis the mass matrix for $(\phi,h)$ is
\begin{equation}
\left( \begin{array}{cc}  
m^2_r + 2 \gamma^2 \lambda v^2  & 2 \gamma \lambda v^2 
 \\
 2 \gamma \lambda v^2  & 
2 \lambda v^2
\end{array} \right) \hbox{ .}
\end{equation}
In this basis, however, the kinetic terms are not diagonal,
but are given by
\beq
{\cal L}_{kin}=\frac{1}{2} (1+\gamma^2) (\partial \phi )^2+\frac{1}{2}
(\partial h)^2+\gamma \partial \phi \partial h.
\eeq
The kinetic mixing is undone by the field redefinition\footnote{
Note, that the radion also has kinetic mixing with the graviton. 
Due to the weak dependence of $M_{Pl}$ on
$b$, however, this mixing is proportional to $\Omega ^2 _0$, and
is thus negligible compared to the Higgs mixing considered here.} 
$\phi- \phi _0  =\bar{\phi}$ 
and 
$h=\bar{h} - \gamma \bar{\phi}$.
In the $(\bar{\phi},\bar{h})$ basis the mass matrix is also 
diagonalized. So $(\bar{\phi},\bar{h})$ are also mass eigenstates, 
with mass squared $m^2_r$ and $2 \lambda v^2$, respectively 
\footnote{This is also seen if in Eq. (\ref{SMl1}) 
we rescale the fields by the vev of 
$\Omega_b$ instead of the full quantum $\Omega_b$.}. 
Note that these are unchanged from their naive expectation.
The relation between the gauge and physical basis is then 
\beq
\phi =\phi _0 + \bar{\phi}
\eeq
and 
\beq
h=\bar{h} - \gamma \bar{\phi} \hbox{ .}
\label{hrmixing}
\eeq

Thus the couplings of the radion to SM fields are similar to the neutral 
Higgs, and may be important for collider phenomenology 
since $\gamma$ may not be small.
We point out that the linear couplings of the radion to SM fields 
may also be 
obtained from : 
\beq
\bar{\phi} \frac{ \delta S_{SM}}{ \delta \phi} \large | _{\phi _0} 
= \bar{\phi} \frac{ \delta S_{SM}}{ \delta \tilde{g}^{\mu \nu}} 
\frac{ \delta \tilde{g}^{\mu \nu}}{ \delta \phi}= - \bar{\phi}
\sqrt{ \tilde{g}} 
\tilde{T}_{\mu \nu} \tilde{g}^{\mu \nu} \frac{\Omega ' _{\phi}} 
{ \Omega _{\phi}}
= \gamma \frac{ \bar{\phi} }{v } T \hbox{ .}
\eeq
So to $O(\bar{ \phi })$ the radion couples to the trace of the stress--tensor.
To this order, this is equivalent to the coupling of the radion to 
the SM in the 
large extra dimension scenario, except that here the suppression is 
TeV rather than $M_{Pl}$.

This discussion then  
demonstrates that 
at the EW scale the theory contains two 
neutral, CP even, spin--0 
particles coupled to the SM fermions and gauge bosons. 
The tree--level mass parameters are the same, however, as in the 
SM, since the vev of the radion does not break 
electroweak symmetry. 
After EWSB the radion has both mass and kinetic 
mixing with the neutral Higgs, which after diagonalization 
results in  
renormalizable interactions between the physical radion and the 
SM fields. These interactions are scaled by $v/(\sqrt{6} \Lambda_W)$, 
so  
the radion is  more weakly (or more strongly!)
coupled to SM fields than the neutral Higgs.
We also point out that the radion also couples to the 
kinetic energy terms, and these operators become important 
at EW energies. 
It is then interesting 
to determine the current 
experimental limits on the radion mass, and 
to what extent future 
collider experiments can distinguish between 
the radion and neutral Higgs boson. We note that these important 
experimental  
issues are completely determined by the 
precise value of $\Omega_0$.

The dominant decay of the radion is to $\bar{t}t$, $ZZ$ or $WW$ if 
kinematically allowed, otherwise to $\bar{b} b$. 
At $e^+e^-$ colliders it can be produced in association 
with $Z$, so its collider signatures are similar to the neutral 
Higgs, but with suppression or enhancement of the rate. 
The Higgs may also decay to two radions (if kinematically allowed), 
and may be competitive with its SM decay modes.

The cosmological implications of this is that the 
radion can decay to SM particles before the 
start of BBN. Of course, a requirement
on early
cosmology for this to occur 
is that by 
the start of BBN the radion was close to its present--day 
value. Then the lifetime of the radion was small enough, and  
the energy density in coherent oscillations of the radion was 
be transfered to radiation by the start of BBN. 

We also note that if matter is present on the Planck brane, 
the coupling of the radion to that brane is $O(1/M_{Pl})$ or 
smaller (see Appendix E). 
Consequently, 
the radion has a very small branching fraction to decay on the 
Planck brane. 
This means that in the early universe the coherent oscillations 
of the radion 
would predominantly 
dump their energy into the TeV brane, and only transfer a 
small fraction $O(($TeV$/M_{Pl}))^2$ to the Planck brane. 
It also implies 
that during the 
time that $b$ was close to its present--day value, matter on the 
Planck brane was not in thermal equilibrium with the matter on the 
TeV brane.

We conclude this section by noting, that (just as for the case of the 
KK gravitons) the radion of the RS model has very different properties 
from the radion of the large extra dimensions of~\cite{Nima}.
In the case of large extra dimensions, the lower bound on the mass
of the radion is $10^{-3}$ eV from  the gravitational force 
measurements. In these models there is also an upper bound on the radion mass,
which is obtained by requiring that the curvature radius is smaller
than the physical size of the extra dimensions. This upper bound 
varies between $10^{-2}$ eV for two extra dimensions to
$20$ MeV for six extra dimensions. Thus in these models the
radion is typically very light, with mass between $10^{-3}$ eV and 
$20$ MeV, while its couplings to the SM fields are suppressed by
$M_{Pl}$. In the RS model however, the radion has unsuppressed
couplings to the SM fields, which result in a lower bound of order
$10 - 100$ GeV for the radion mass, while there is no upper bound from the
curvature constraint. Thus one can see, that the radion in these two models
has very different properties, and they should be easy to distinguish
experimentally.

\section{Dark Matter on the Planck Brane or in the Bulk?}
\setcounter{equation}{0}
\setcounter{footnote}{0}
\label{section6}


The expansion rate of the universe is driven by three sources: matter living 
on the TeV brane, matter living on the Planck brane, and matter living 
in the bulk:
\beq
H^2= \frac{\dot{a}^2}{a^2 }= 
\frac{8 \pi G_N} {3}
\left(\rho_* +\rho \Omega^4 _0 + \rho_{bulk} \right)   \hbox{ ,}
\label{dark}
\eeq 
where $\rho_{bulk}=b_0\int_{-\frac{1}{2}}^{\frac{1}{2}} \Omega (y)^4 
\rho_{bulk}(y) dy $, with $\rho_{bulk}(y)$ being the five-dimensional
bulk energy density.
(Recall that for late--cosmology 
it is a good approximation to set
$V_r=0$ and $\dot{b}=0$ in the Hubble law. Also   
the time 
dependence in $b$ due to the shift in $b$ contributes $O(\rho^2)$ to 
$H^2$ which is negligible for late--cosmology).

We see from Eq. (\ref{dark}) that the appearance of $\rho_*$ and 
$\rho_{bulk}$ in the expansion rate is potentially dangerous. This 
is because the natural scale on the Planck brane is $O(M^4_{Pl})$, 
and so this energy density could easily overclose the universe. 
But we will argue below that natural mechanisms 
exist to suppress the energy density on the Planck wall  well--below 
its natural value. But of course a careful computation within a 
specific model for early cosmology is required to establish that 
these suppressions are indeed suffucient.  

Since the natural scale on the Planck brane is $M^4_{Pl}$, we see 
from the Hubble Law that 
there are two different energy scales appearing in the effective 4D theory,
which is at first surprising. This suggests 
that observers living on the TeV brane will measure
Planck--mass particles living on the Planck brane 
to really have Planckian masses, even 
though the cutoff on the TeV brane is $O($TeV$)$. In fact, this is 
confirmed in Appendix D by a computation of the Newtonian force 
law between a particle on the Planck brane and a particle on the TeV brane. 
We find that the Newtonian force law between particles of bare mass 
$m_1$ and $m_2$ located at $y_1$ and $y_2$ along the extra dimension 
is given by
\beq
F_N = G_N m_1 \Omega(y_1)m_2 \Omega(y_2) /r \hbox{ .}
\eeq
So the Newtonian force law and Hubble Law for matter living on different 
branes are consistent. 

Thus one can see from (\ref{dark}), that there are two new candidates 
for dark matter: matter in the bulk and matter on the Planck brane.
The radion is not a candidate for dark matter, since as we have 
seen in the previous section, it has a short lifetime (of the 
order $M_{weak}^{-1}$), and is therefore not abundant today.

Matter in the bulk can naturally be dark matter, since its natural
mass scale is ${\cal O} (TeV)$\footnote{A 5D scalar 
with $O(M_{Pl})$ bare mass appears in the 4D effective theory as a 
Kaluza-Klein tower of scalars. The lightest mode has a 
mass of $O($TeV$)$\protect\cite{Wise1}.}, and it has interactions
with the SM fields suppressed by the TeV scale. Thus it could have been
in thermal equilibrium with the TeV brane when the temperature of the
Universe was ${\cal O}(TeV)$. If these bulk fields are unstable,
then they will predominantly decay to SM fields, rather than matter 
living on the Planck brane, since there the couplings are suppressed by
$M_{Pl}$. However, if those bulk fields are stable, and since it is 
natural for  
their couplings
to SM particles to be TeV suppressed 
(see Appendix E), their annihilation cross section
may turn out to lead to interesting relic densities today, 
which may serve as 
dark matter. 

Another interesting possibility is to have matter on the Planck brane serve
as dark matter. The natural mass scale for particles on the 
Planck brane is either $M_{Pl}$ or 0 (massless). 
In either case one might worry that matter on the Planck 
brane would overclose the Universe. However, matter there is very 
weakly coupled to bulk fields, since its couplings are 
suppressed due to the small wavefunction of the bulk fields 
at the Planck brane, in addition to the suppression of 
$M_{Pl}$ for non--renormalizable operators. 
We illustrate this in Appendix E,
where we calculate the couplings of bulk scalars to the Planck (and the TeV)
branes. Therefore matter on the Planck brane has  
presumably never been in thermal equilibrium with our brane after
inflation. 
Even if massive matter there was in a thermal equilibrium 
with the TeV brane (or if the temperatures on the two branes were 
once comparable), its density is suppressed by a Boltzmann factor, 
and is therefore greatly suppressed. Such a suppression would however
not happen for massless fields on the Planck brane
(unless the temperature itself on the Planck brane is 
for some reason much smaller than on
the TeV brane), thus 
the presence of such massless fields is strongly disfavored, which
is a constraint for model building. It would be 
worthwhile to investigate whether 
non-thermal production of Planck brane matter 
could result in interesting
relic densities \cite{CK} and thus serve as dark matter.



\section{Conclusions}
\setcounter{equation}{0}
\setcounter{footnote}{0}
\label{section7}

In this paper we have considered the effect of a stabilization mechanism 
for the radion on the cosmology of the RS model. We have found that the 
previously discovered unconventional cosmologies result from 
the fact that the radion was not stabilized in the basic RS scenario.
The constraint between the matter on the two 
branes was a consequence of trying to find a static solution 
to the radion equations of motion without actually stabilizing it.
Once a stabilization mechanism is added, the constraint disappears,
and the ordinary four 
dimensional FRW Universe is recovered at low temperatures 
if the radion is stabilized with a 
mass  of the order of
its natural scale $M_{weak}$.

In this paper we have also pointed out that after electroweak
symmetry breaking the radion has renormalizable couplings 
to matter on the TeV brane. 
These interactions are uniquely determined by the precise 
value of the warp factor. The production and detection of the radion 
are then similar to the detection of the SM Higgs, but with enhanced
or suppressed rates. An important 
consequence of these interactions for cosmology is that there 
is no radion moduli problem, as the radion decays into SM particles 
long before the start of BBN. 
We have also speculated that matter in the bulk or on the Planck brane
are new candidates for dark matter.

\section*{Note Added}
While this paper was being concluded, several papers related to this subject
appeared~\cite{Japanese,Tye,BDEL}, however none of these papers consider the
central issue of our paper, the effect of radion stabilization on the 
cosmology of the negative tension brane. 
Ref.~\cite{Japanese} derives the four dimensional gravity equations in 
a covariant formalism, and concludes similarly to \cite{CGKT,CGS}: the
expansion of the positive tension brane is conventional, however, there is 
a sign difference for the case of the negative tension brane. 
Refs.~\cite{Tye,BDEL} consider the cosmology of the positive tension brane.
Ref.~\cite{Tye} concludes that after inflation 
one can always end up with the ``correct'' 
expanding solution found in \cite{CGKT}. 
Ref.~\cite{BDEL} finds the exact solutions
in the bulk to the Einstein equations with matter on the branes and 
background cosmological constants. The solution presented in Appendix A of
this paper is a perturbative expansion of the exact solution of~\cite{BDEL},
obtained independently from~\cite{BDEL}.

\section*{Acknowledgments}

We are extremely grateful to Riccardo Rattazzi for several detailed 
conversations and suggestions, and to 
Andy Cohen, Ann Nelson, John--March Russell and 
Raman Sundrum for helping provide the right intuition. 
We also thank Nima Arkani-Hamed, 
Pierre Binetruy, Kiwoon Choi, 
Michael Dine, Howie Haber, Nemanja Kaloper, Chris Kolda, 
Anupam Singh and Mark Wise for useful discussions.
C.C. and L.R. 
thank ITP at UC Santa Barbara 
for their hospitality where part of this work was 
completed. C.C. and L.R. also thank the Aspen Center for Physics for its
hospitality during the early phase of this work. M.G. thanks the T-8 group
at Los Alamos National Laboratory for its hospitality while this paper was
completed. C.C. is a J. Robert Oppenheimer fellow at the Los Alamos 
National Laboratory. The research of C.C. is supported by
the Department of Energy under contract W-7405-ENG-36.
The work of M.G. was supported in part by the Department 
of Energy. The work of L.R. was supported in part by the Department of
Energy under cooperative agreement DE-FC02-94ER40818 and under
grant number DE-FG02-91ER4071. The work of J.T. is supported in part 
by the NSF under grant PHY-98-02709.
The work of C.C. and L.R. while 
at the ITP in Santa Barbara has been supported by the NSF under grant 
PHY-94-07194.

\appendix

\section*{Appendix}

\section{The Solution without a Stabilized Radius}
\setcounter{equation}{0}
\setcounter{footnote}{0}
\label{appendixA}
In this Appendix we find an approximate solution to Einstein's equations
in the RS model with matter on the branes for the case when the radius
is not stabilized. The main purpose of this
is to illustrate the generic feature of all such solutions of brane 
models without a stable radius: there appears to be a
constraint between matter on the two branes.
This is similar to the relations found in
\cite{BDL,kaloper,KimKim}.
It has been suggested \cite{BDL} that
such relations are of topological origin and are very generic to
brane world scenarios. 
In the Sections \ref{section3} and 
\ref{section4} we argue that the real origin of this 
constraint is the  
requirement of a stable 
radius even in the absence of a stabilizing potential. 
Otherwise the extra dimension itself 
will expand or contract. The 
presence of the stabilizing potential then prevents the expansion 
of the extra dimension, and a phenomenologically acceptable cosmology 
follows. 

In order to find a solution we
assume the form of the metric to be
\beq 
ds^2&=&n(y,t)^2 dt^2-a(y,t)^2 (dx_1^2+dx_2^2+dx_3^2)-b^2 dy^2,
\eeq
that is we assume we have a time-independent constant radion $b$ even though
there is no stabilization mechanism. We also assume that 
the matter density on the branes is small compared to the brane tensions
$\rho, \rho_* \ll V$. Then one can find a solution as a perturbative series 
in $\rho$ around the RS solution:
\begin{eqnarray}
&& a(y,t)=a_0(t) e^{-|y|b_0 m_0} (1+\sum_{l=1}^{\infty} \rho_*(t)^l f_l(y)),
\nonumber \\
&& n(y,t)= e^{-|y|b_0 m_0} (1+\sum_{l=1}^{\infty} \rho_*(t)^l g_l(y)).
\end{eqnarray}
One could start with a more general ansatz, where instead of explicitly 
expanding in $\rho_* (t)$ we expand in arbitrary functions of time. However,
the ``jump equations'' discussed below will immediately tell us that this 
function of time is proportional to $\rho_* (t)$.

Since $\rho_* \ll V$, we will keep only terms that are linear in $\rho_*$,
thus our ansatz for the solution will be 
\begin{eqnarray}
&& a(y,t)=a_0(t) e^{-|y|b_0 m_0} \left( 1+\rho_*(t)f(y)\right),
\nonumber \\
&& n(y,t)= e^{-|y|b_0 m_0} \left( 1+\rho_*(t) g(y)\right).
\end{eqnarray}
We know from \cite{BDL}, that the jumps of $a$ and $n$ are related to 
the density and pressure in the following way:
\begin{eqnarray}
\frac{[a'(0,t)]}{a(0,t)b_0}=-\frac{\kappa^2}{3} 
\left(V+\rho_*(t) \right) \hbox{ ,} \nonumber \\
\frac{[n'(0,t)]}{n(0,t)b_0}=\frac{\kappa^2}{3} 
\left(-V+\rho_*(t)(2+3w_*)\right), 
\end{eqnarray}
where $[h]$ denotes the jump in the function $[h(0,t)]=(h(0+\epsilon)-
h(0-\epsilon))$. From these equations we learn what the jumps of $f$ and $g$
have to be:
\begin{eqnarray}
&&[f'(0)]=-\frac{\kappa^2 b_0}{3}, \nonumber \\
&& [g'(0)]=\kappa^2 b_0 (w_*+\frac{2}{3}).
\end{eqnarray}
The $G_{05}$ equation 
tells us that $f(y)$ and $g(y)$ have to be proportional to 
each other (up to a constant):
\beq
\dot{\rho_*}+3 \frac{\dot{a_0}}{a_0}\rho_* \left( 1+w_*\frac{g'(y)}{f'(y)}
\right)=0,
\eeq
which implies that $g'(y)/f'(y)=const.$, and using the values of the jumps 
of the derivatives together with the fact that we have an $S^1/Z_2$ orbifold 
implies that the $G_{05}$ equation 
just reduces to the usual conservation of energy 
equation of the form
\beq
\dot{\rho_*}+3 \frac{\dot{a}_{0}}{a_{0}} (\rho_* +p_*)=0 \hbox{ .}
\label{energy1}
\eeq
Similarly, at the other brane we have 
\beq
\dot{\rho}+3 \frac{\dot{a}_{1/2}}{a_{1/2}} (\rho +p)=0 \hbox{ .}
\label{energy2}
\eeq
Plugging the ansatz into the other equations then implies that
\beq 
f(y)=-\frac{\kappa^2}{12 m_0} (e^{2|y|m_0 b_0}-1), \nonumber \\
g(y)=\frac{\kappa^2 (2+3w_*)}{12 m_0} (e^{2|y|m_0 b_0}-1).
\label{correction}
\eeq
The constant in $f$ and $g$ have been determined such, that in the limit
$m_0\to 0$ the solution exactly reproduces the bulk solution presented 
in \cite{BDL}. In order to satisfy Einstein's equations 
the function $a_0(t)$ has to satisfy the following Friedmann-type 
equations:
\begin{eqnarray}
&& \left( \frac{\dot{a_0}}{a_0} \right)^2=\frac{1}{3} \kappa^2 m_0 \rho_* ,
\nonumber \\
&& \left( \frac{\dot{a_0}}{a_0} \right)^2+\frac{\ddot{a_0}}{a_0}=
\frac{\kappa^2 m_0}{6} (\rho_* -3p_*).
\label{exp}
\end{eqnarray}
From (\ref{correction}) 
one can see that the Hubble constant at different points
along the extra dimension is given by
\beq
\left(\frac{\dot{a}(y)}{a(y)}\right)^2 \equiv 
H^2(y)=H^2(0) \left( 1+\frac{\kappa^2}{2m_0} (e^{2|y|m_0b_0}-1)(\rho_*+p_*)
+{\cal O}(\rho^2 _*)\right),
\eeq
which for $\rho_* \ll M_{Pl}^4$ is a very slowly varying function of $y$.
This can be seen by comparing $H^2(\frac{1}{2})$ and $H^2(0)$:
\beq
H^2(\frac{1}{2})-H^2(0)\sim  H^2(0) 
\frac{\kappa^2e^{m_0b_0}}{2m_0} (\rho_*+p_*).
\eeq
If $e^{m_0b_0}\sim  10^{30}$, and $\frac{\kappa^2}{m_0}\sim  
\frac{1}{M_{pl}^4}$, then $H^2(\frac{1}{2})-H^2(0)\sim 
H^2(0) \frac{\rho_* (1+w)}{(10^{10}\, {\rm GeV})^4}$.
Thus for temperatures below $10^{10}$ GeV on the Planck brane the two branes
will expand together, but the expansion rate is completely fixed by the 
matter on the Planck brane. This shows, that for these temperatures 
there should be a sensible effective four dimensional theory 
describing the evolution of the Universe, since $a(0)\sim  a(\frac{1}{2})
\sim  \langle a \rangle = \int_{-\frac{1}{2}}^{\frac{1}{2}} a(y) dy.$
Below we will show that the effective theory calculations do indeed 
reproduce (\ref{exp}).

  From the point of view of the 
negative tension brane one can understand the previous Hubble Law
from the following consideration.
The above solution depends only on $\rho_*$ and $p_*$, and not 
on $\rho$ or $p$. This means then that the jump equation at $y=1/2$
completely determines the energy density on the negative tension brane. 
One finds that the condition is simply given by
\begin{equation}
\rho_*(t)=-\rho(t) \Omega^2_0, \ \ p_*(t)=-p(t) \Omega^2_0.
\label{c1}
\end{equation}
This explains the results found in \cite{CGKT,CGS}\footnote{Note, that a 
factor of $\Omega_0^2$ has been erroneously omitted in (14) of 
\protect\cite{CGKT} and in (5) of \protect\cite{CGS} in the case of 
$\rho_-$. The reason for this extra factor is that from the point
of view of the negative tension brane the fundamental scale of gravity is
of the order of $M_{Pl}^2 \Omega_0^{2} \sim 1$ TeV.}:
the expansion equation has the wrong sign if one assumes $\rho >0$. 
In \cite{CGKT} $\rho >0$ was 
assumed in order to get a sensible phenomenology 
on the TeV brane. However, we see that the constraint (\ref{c1}) implies 
that for $\rho >0$ one needs to have $\rho_*<0$. But from
(\ref{exp}) we see that $\rho_*$ cannot be negative. Therefore, in this
case $\rho$ on our brane must be negative. This is completely due to the
constraint (\ref{c1}), and not a consequence of the breakdown of the
effective 4D theory.

One can check this result by explicitly calculating the effective 
4D action for this setup, using the insight from the solution.
The full 5D action is given by
\beq
-\int d^5x \sqrt{g} (\frac{R}{2\kappa^2}+\Lambda )
+\int d^4 x\sqrt{g^{ind}_0} {\cal L}_{Pl} +\int d^4x\sqrt{g^{ind}_{1/2}} 
{\cal L}_{TeV}.
\eeq
To calculate the effective 4D action one needs to integrate over the
extra dimension. We do this substituting for the metric the ansatz
\begin{eqnarray}
a(t,y)=a_0(t)
(1-\frac{\kappa^2 \rho_*}{12 m_0} (e^{2|y|m_0b_0}-1)), \nonumber \\
n(t,y)=(1+\frac{\kappa^2 \rho_*(2+3w_*)}{12 m_0} (e^{2|y|m_0b_0}-1)),
\end{eqnarray}
however without assuming that $a_0$ and $\rho$ satisfy the above
Friedmann equations.
The result of integrating over the fifth coordinate $y$ to linear order
in $\rho$ and $\rho_*$ is
\begin{eqnarray}
S_{eff}=&& 
\int dt a_0(t)^3 \frac{3 (1-\Omega^2 _0)}{\kappa^2 m_0}
\left( \left(\frac{\dot{a_0}}{a_0}\right)^2
+\frac{\ddot{a_0}}{a_0}\right) + \int dt a_0(t)^3 {\cal L}_{Pl}+
\int dt a_0(t)^3 \Omega_0^4 {\cal L}_{TeV}
\nonumber \\
=&& \int dt  a_0^3(t)\left(
-\frac{1}{2} 
M_{Pl}^2 {\cal R}^{(4)}+ {\cal L}_{Pl}+ \Omega_0^4 {\cal L}_{TeV}
\right) .
\end{eqnarray}
This is exactly what we expect: the action for the expanding universe with 
the correct Planck constant, and with total energy density 
$\rho_{total}= \rho_* + \rho \Omega^4 _0$. 
As expected, the energy density on the 
''TeV'' brane contributes $ \rho \Omega^4 _0$ to the expansion 
rate.
The expansion rate 
obtained from this effective action also agrees with the expansion 
rate for the induced metric in the five--dimensional theory. In fact, 
the effective theory gives  
\begin{equation}
H^2 =\frac{8 \pi G_N} {3} \left(\rho_* + \rho \Omega^4 _0 \right) \hbox{ ,} 
\end{equation} 
which after substituting  
the constraint Eq.~(\ref{c1}), and the relation between $M_{Pl}$ and 
the five-dimensional parameters, Eq.~(\ref{mpl}), identically agrees with 
the 
expansion rate Eq.~(\ref{exp}) in the five--dimensional theory.

Thus we can see that the effective theory picture is as expected in agreement
with the detailed form of the solutions to the Einstein equations:
there is no breakdown of the effective theory picture since the two branes
expand together. The 
only puzzle that remains is why one needs to have a relation between the
matter fields on the two branes of the sort in Eq.~(\ref{c1}).
As we will see in Sections 2 and 3, this is a consequence of the
fact that we have not stabilized the radius.
It simply tells us, that if we put in matter that does not
obey this constraint the extra dimension will want to expand itself.
However, if the extra dimension is stabilized, no such constraint should
exist and the cosmological expansion is given by the conventional Friedmann
equations in the 4D effective theory with no constraint on the magnitude of
the matter one can add to the system.

\section{The Case of Vanishing Cosmological Constants}
\setcounter{equation}{0}
\setcounter{footnote}{0}
\label{appendix}

In order to show that our results are generic for any brane model,
we will convince ourselves by using arguments similar to
those in Section 4 that the ordinary Friedmann equations are recovered
for the case of vanishing background cosmological constants as well
(the case considered in \cite{BDL}), once
the radion is stabilized. To see this in detail, we again look for the
solutions to the $00$, $ii$ and $05$ components of the Einstein
equations only, and just like in Sec. \ref{section3} we do not
require that that the $55$ component is satisfied, since this will be solved
by adjusting the radion, which we once again assume to be very heavy. 
We assume that we include matter density $\rho (t)$ on ``our brane'', and
matter density $\rho_* (t)$ on the other brane. (Note, that since the
radius is stabilized, the presence of a second brane in the case of 
vanishing background cosmological constants is not even necessary. One
can obtain all the relevant formulae for this case by just setting
$\rho_*=0$.) We look for an approximate 
solution (i.e. valid to ${\cal O}(\rho_*, \rho)$)
on the $S^1/Z_2$ orbifold in the form
\begin{eqnarray}
&& a(t,y)= a_0(t) (1+\alpha \rho_* (t) (y-\frac{1}{2})^2+\beta \rho (t) y^2),
\nonumber \\
&& n(t,y)= (1+\gamma \rho_* (t) (y-\frac{1}{2})^2+\lambda \rho (t) y^2),
\nonumber \\
&& b(t,y)=b_0(1+ \delta b),
\end{eqnarray}
for $0\leq y \leq \frac{1}{2}$, and the solutions for the other regions 
are obtained by reflecting around $y=0$ or $y=\frac{1}{2}$. 
In this ansatz $\alpha, \beta, \gamma$ and $\lambda$ are constants. From the 
jump equations we obtain that 
\beq \alpha=\beta= \frac{\kappa^2 b}{6}, \ \gamma= -\frac{(2+3w) 
\kappa^2 b}{6}, \ \lambda= -\frac{(2+3w_*) \kappa^2 b}{6},
\eeq
where $p=w \rho$ and $p_*=w_*\rho_*$. One can easily check that with this 
choice of constants the relevant components of the Einstein equation are 
satisfied  to first order in $\kappa^2 b_0 \rho$ which is assumed to be 
small, and $a_0$ satisfies the following Friedmann equations:
\begin{eqnarray}
&& \left( \frac{\dot{a}_0}{a_0}\right)^2 
=\frac{\kappa^2}{3b_0} (\rho +\rho_*),
\nonumber \\
&& \left( \frac{\dot{a}_0}{a_0} \right)^2 + 2 \frac{\ddot{a}_0}{a_0}=
-\frac{\kappa^2}{b_0} (w\rho +w_* \rho_*).
\end{eqnarray}
These are the correct 4D FRW equations with the correct 
normalization, since $H^2 \propto \rho$, rather than 
$H^2 \propto \rho^2$, and 
as  $M^2_{Pl} =b_0 / \kappa^2$ is 
the relation between the 4D and 5D parameters. The size of $\delta b$ is 
obtained from inspecting the $G_{55}$ equation and is readily seen to be
$O(\rho^2, \rho_* ^2, \rho \rho_*)$. 
Once again, we find that after stabilization of the radion the ordinary 
four dimensional Friedmann equations are recovered, with no constraint on
what kind of matter one can include on the branes. Thus the ordinary FRW 
cosmology is recovered which results in ordinary BBN.

\section{Radion Mass in the Goldberger-Wise Stabilization Mechanism}
\setcounter{equation}{0}
\setcounter{footnote}{0}
\label{GWA}
Here we compute the radion mass using the Goldberger--Wise (GW)
mechanism for radion stabilization \cite{GW}. It is found that 
the radion is naturally of $O($TeV$)$. 

In their mechanism for generating a stabilizing potential, 
a bulk scalar field $\Phi$ is introduced which has a 5D action 
\beq
S_{\Phi} = \frac{1}{2} 
\int d^5 x \sqrt{\tilde{G}} \left( 
\tilde{G}^{AB} \partial _A \Phi \partial _B \Phi -m^2_S \Phi^2 \right) 
\hbox{ ,} 
\eeq
where $\tilde{G}_{AB} $ is the full 5D metric which includes the 
RS warp factor. 
The bulk mass $m^2_S$ is assumed to be somewhat 
smaller than the Planck scale. So 
\beq
\epsilon' = \frac{4 m^2_{S}}{m^2_0}  
\eeq
is a small quantity. 
The bulk scalar also contains two potentials on the Planck and 
TeV brane, 
\beq
S_{Pl} = \int d^4 x \sqrt{-g} \lambda _h (\Phi^2 -v_h^2)^2 
\eeq
and 
\beq
S_{TeV} =\int d^4 x \sqrt{-g} \lambda _v (\Phi^2 -v_v^2)^2 \hbox{ .}
\eeq
Note that $v_v$ and $v_h$ have mass dimension $3/2$.
GW solve the equation of motion for $\Phi$, and find that 
it has a non--trivial dependence in the bulk. 
In the large $\lambda_{v,h}$ limit, GW find that 
$\Phi(0)=v_h$ and $\Phi(1/2)=v_h$, so that in this 
limit the brane--potentials do not contribute to the 4D energy. 
However, $\Phi$ contains both potential and (gradient) kinetic 
energy in the bulk, so that the energy density in the bulk is 
inhomogeneous. 
The radion potential 
is then obtained by substituting the 
classical solution for $\Phi$ into 
the above action and integrating over the extra dimension.
GW find that the resulting  
4D effective potential for the radion is given by  
\beq
V_r(b)=4 m_0 e^{-2 m_0 b} \left(v_v-v_h e^{-\epsilon' m_0 b/2} \right)^2
(1+ \frac{\epsilon'}{4} ) 
- \epsilon' m_0 v_h e^{-(4+ \epsilon')m_0 b/2} 
\left( 2 v_v -v_h e^{-\epsilon' m_0 b/2} \right)
 \hbox{ .}
\eeq
In their computation GW dropped terms of $O({\epsilon' }^2)$ in 
$V_r$, so the following computation of the radion mass 
is only accurate to this order. 
Then 
\begin{eqnarray}
\frac{V'_r(b)}{4 m^2_0 e^{-2 m_0 b}} &=&
-\left(v_v-v_h e^{-\epsilon' m_0 b/2}\right)
\left(2(v_v-v_h e^{-\epsilon' m_0 b/2}) + 
\frac{\epsilon'}{2} (v_v -3 v_h e^{-\epsilon' m_0 b/2}) \right)
\nonumber \\
& &
+ \frac{\epsilon'}{2} v_h e^{-\epsilon' m_0 b/2} \left(2 v_v 
-v_h e^{-\epsilon' m_0 b/2} \right) +O({\epsilon'} ^2) 
\end{eqnarray}
There is one trivial solution at $b \rightarrow \infty$. 
In addition, we expect a local maximum and a local minimum.
These 
two 
are at 
\beq
v_v- v_h e^{-\epsilon' m_0 b/2}= \delta 
\eeq
where $\delta$ is the solution to 
\beq
\delta^2 - \frac{\epsilon'}{1+\epsilon'} \frac{v^2 _v}{4} 
- \frac{\epsilon'}{1+\epsilon'}\frac{v_v}{2} \delta = 0  \hbox{ .}
\eeq
That is 
\beq
\frac{\delta}{v_v}= \frac{1}{4}\frac{\epsilon'}{1+ \epsilon'}
\pm \frac{\sqrt{\epsilon '}}{2(1+\epsilon')} \left(1+ 
\frac{5}{4} \epsilon' \right)^{1/2} \hbox{ .}
\eeq
Then at these two extremum 
\beq
V''(b_0)=-8 \epsilon' \delta 
 m^3_0 v_v   e^{-2 m_0 b_0} + O({\epsilon'}^2) 
\hbox{ .} 
\eeq
Since this is $O(\delta \epsilon')$, we keep $\delta$ to 
$O(\sqrt{\epsilon'})$. Using the above solution for $\delta$, we find 
that to this order
\beq
\delta^{\pm} = \pm \frac{\sqrt{\epsilon'}}{2}  v_v  \hbox{ .}
\eeq
So $\delta ^{-}$ is the solution corresponding to the local 
minimum. This is expected, since there should be 
local maximum 
between this local minimum and the minimum at $b \rightarrow 
\infty$. Then 
\beq
V''(b)= 4 {\epsilon'}^{3/2}  m^3_0 v^2 _v e^{-2 m_0 b_0} + O({\epsilon'}^2) 
\hbox{ .} 
\eeq
Using 
Eq.~(\ref{radionmass}) and $\Omega_0 =e^{-m_0 b_0/2}$, the radion 
mass is 
\begin{eqnarray}
m^2_r &=& \frac{8}{3} 
{\epsilon'}^{3/2} \frac{m_0 v^2 _v}{M^2_{Pl}}  \Omega^2 _0 
\nonumber \\
 &\sim&  {\epsilon'}^{3/2} \Lambda^2 _W \hbox{ ,}
\end{eqnarray}
since $m_0 \sim M_{Pl}$ and $v_v \sim M^{3/2}_{Pl}$. 
So the radion mass is $O($TeV$)^2$. For the GW solution, 
$\epsilon' \sim  1/40$, so the radion is roughly an order of 
magnitude below $\Lambda_W$.

\section{Newton Force Law Between Particles on the TeV and Planck Branes}
\setcounter{equation}{0}
\setcounter{footnote}{0}
\label{Newton}
We have seen that the expansion in the effective four dimensional
theory is given by
\beq
H^2= \frac{\dot{a}^2_0}{a^2 _0}= 
\frac{\kappa^2 m_0} {3}\frac{1}{1-\Omega^2_0}
\left(\rho_* +\rho \Omega^4 _0 + \rho_{bulk} \right)   \hbox{ .}
\eeq 
That two mass scales, $O(M_{Pl})$ in $\rho_*$, and 
$O($TeV$)$ in $\rho \Omega^4_0$, appear in the expansion rate 
for the effective theory is at first surprising. This suggests 
that observers living on the TeV brane will measure
''Planck--massed'' particles living on the Planck brane 
to really have ''Planckian'' masses, even 
though the cutoff on the TeV brane is $O($TeV$)$.  
In fact, this is confirmed by a computation of the Newtonian force 
law between a particle on the Planck brane and a particle on the TeV brane. 
 To see this, the 5D Lagrangian 
for a point mass 
particle with bare mass $m_1$ living 
on the Planck brane, and a point mass particle 
with bare mass $m_2$ living on the TeV brane is 
\beq 
-m_1 \int d^5 x  \delta (y) \delta^4 (x-z_1(\tau)) 
\sqrt{\bar{g} ^{\mu \nu } \dot{x}_{\mu} \dot{x}_{\nu} } 
-m_2 \int d^5 x \delta (y-1/2)   
\delta^4 (x-z_2(\tau)) 
\sqrt{ \bar{g}^{\mu \nu} \dot{x}_{\mu} \dot{x}_{\nu} } \hbox{ ,}
\eeq
where $\bar{g}_{\mu \nu}= \Omega^2(y) g_{\mu \nu}$ and indices 
are raised and lowered with respect to $\bar{g}$.
The stress-energy tensor (with respect to $\bar{g}$)
for one of these particles is 
\beq
\bar{T}_{\mu \nu} 
= -m_i \frac{1}{\sqrt{\bar{g}}} \delta(y-y_i) \delta^4 (x-z_i(\tau))
\frac{\dot{x}_{\mu} \dot{x}_{\nu}}{  \sqrt{ \bar{g}^{\mu \nu} \dot{x}_{\mu} 
\dot{x}_{\nu}}} 
\equiv m_i S_{\mu \nu} ( \bar{g}_{\mu \nu}) \hbox{ .}
\eeq 
But if we express $S$ in 
terms of $g$, and raise and lower indices with $g$, then 
\beq
\bar{T}_{\mu \nu}= m_i \Omega^{-1}(y_i) S_{\mu \nu}( g_{\mu \nu}) .
\eeq
But with $\bar{g}= \Omega^2 + \bar{h}$, 
the (linearized) interaction with gravity is 
\beq 
\int dy \sqrt{\bar{g}} \bar{T}_{\mu \nu} \bar{h}^{\mu \nu}
= m_i \int dy \Omega^3 \sqrt{g} S_{\mu \nu} \bar{h}^{\mu \nu} \hbox{ .} 
\eeq
This should be expressed in terms of the 
graviton zero mode, which is 
$h_{\mu \nu}(x,y)= h_{\mu \nu}(x)$ where 
$\bar{h}_ {\mu \nu}(x,y) 
= \Omega ^{2}(y) h_ {\mu \nu}(x,y)$. Then  
the coupling of the particles to the zero mode is 
\beq
- m_i \int dy  \Omega \sqrt{g} S_{\mu \nu}(x,y) h^{\mu \nu}(x)  \hbox{ .}
\eeq
So we see that each particle couples to $h$ with strength $\Omega(y_i) m_i$. 
The Newtonian force computed from the exchange of the zero mode only, 
$h(x,y)=h(x)$, is then 
\beq
F_N = G_N m_1 \Omega(y_1)m_2 \Omega(y_2) /r \hbox{ .}
\eeq
So the Newtonian force law and Hubble Law for matter living on different 
branes are consistent. 
This result is consistent with \cite{randall}, where the authors
 found that the 
physical masses are given by $\Omega(y_i) m_i$.
Therefore both mass scales appear in the 
4D effective theory.

\section{Couplings of Bulk Matter}
\setcounter{equation}{0}
\setcounter{footnote}{0}
\label{Couplings}

In this Appendix, we show what the natural suppression scale for couplings
of bulk fields are to matter on the Planck or the TeV branes.
Following \cite{Wise1}, we consider a general bulk scalar $\Phi$
(not necessarily the GW field of radius stabilization) with bulk action
\beq
S_{\Phi} = \frac{1}{2} 
\int d^5 x \sqrt{\tilde{G}} \left( 
\tilde{G}^{AB} \partial _A \Phi \partial _B \Phi -m^2 \Phi^2 \right) 
\hbox{ ,} 
\eeq
where $\tilde{G}_{AB} $ is the full 5D metric which includes the 
RS warp factor, and $m$ is the bare 5D mass of this scalar, assumed to be
of the order of the 5D Planck mass $M$. 
We assume that this field $\Phi$ has some
generic couplings with matter on a brane of the form
\beq
\sqrt{g^{ind}} \frac{(\Phi \Phi)^p}{M ^{3p}} 
\frac{{\cal O}^{4+q}}{M ^q},
\eeq
where ${\cal O}^{4+q}$ denotes a composite operator of fields living on
one of the branes, and has mass dimension $4+q$. We assume that in the bare
Lagrangian every suppression scale is proportional to the 
$M$. Performing the usual
field redefinition for the fields living on the brane
results in ${\cal O}^{4+q}\to \Omega^{-q}(y_i) \tilde{{\cal O}}^{4+q}$,
where $\tilde{{\cal O}}^{4+q}$ has canonically normalized fields and 
physical 
masses, and  $\Omega (y_i)$ is the warp factor at the position of the
brane, $\Omega (0)=1$, $\Omega (\frac{1}{2})=10^{-15}$. 
We can see that the effect of this this redefinition is to turn the
suppression factor $M ^q$ for the operators on the brane into 
$\sim \Lambda^q _W$ for matter on the TeV brane. 
For matter on the Planck brane
the suppression factor remains unchanged since no rescaling of the
fields is needed. 

We decompose the field $\Phi$ as \cite{Wise1}
\beq
\Phi (x,y)= \frac{1}{\sqrt{b_0}} \sum_n \Psi_n (x) \varphi_n (y),
\eeq
where \cite{Wise1}
\beq
\varphi_n (y)= \frac{\Omega (y)^{-2}}{N_n}
\left[ J_{\nu} \left( \frac{m_n}{m_0} \Omega (y)^{-1} \right)+
b_{n\nu} Y_{\nu} \left( \frac{m_n}{m_0} \Omega (y)^{-1} \right)\right],
\eeq
$J_{\nu}$ and $Y_{\nu}$ are Bessel functions of order 
$\nu=\sqrt{4+\frac{m^2}{m_0^2}}$, $m_n$ is the mass 
of the Kaluza--Klein mode $\Psi_n$, 
and $N_n$ and $b_{n\nu}$ are normalization
constants. With this substitution for $\Phi$ the interaction of a 
particular KK mode to the brane fields is given by operators 
of the form
\beq
\frac{(\Psi_n \Psi_n)^p}{M ^{3p}} 
\frac{\tilde{{\cal O}}^{4+q}}{(\Omega (y_i)M )^q} (\varphi_n (y_i))^{2p}.
\label{interact}
\eeq
Thus we need to find the approximate value of the wave function of the
KK modes at the position of the two branes. First we calculate the 
suppression
at the TeV brane. We will use the approximate values $\nu \sim 2$, 
$N_n \sim \Omega_0^{-1}$. For small KK masses 
$b_{n\nu}$ is approximately given by 
$\pi m_n^2 /4 m_0^2 \sim \Omega_0^2$, thus for arguments of order one
in the Bessel functions the contribution of $Y_{\nu}$ can be neglected.
Thus the approximate value of $\varphi _n$ is given by 
\footnote{Note that the normalization of these wave functions is 
$\int dy \Omega^2 \phi_{(m)} \phi_{(n)} = \delta_{mn}$.}
\beq
\varphi _n (1/2) \sim \frac{\Omega_0^{-2}}{\Omega_0^{-1}}=
\frac{1}{\Omega_0},
\eeq
which then turns every factor of $M$ in (\ref{interact}) into 
$\sim \Lambda_W$.
Thus the interactions with matter on the TeV wall are suppressed by the
TeV scale as expected.

To find the value of the wave functions close to the origin,
we expand $J_2 (x_n) \sim x_n^2$ and $Y_2 (x_n)\sim -\frac{1}{x_n^2}$ 
for small
values of $x_n=m_n/m_0$. Thus now the contribution of $Y_2$ will dominate,
and we get the approximate value 
\beq
\varphi _n (0) \sim \frac{1}{\Omega_0^{-1}} x_n ^2  \frac{1}{x^2_n}
\sim \Omega_0.
\eeq
Thus the interactions on the Planck brane are suppressed by a scale
higher than $M_{Pl}$, namely $M_{Pl} 10^{15}$. This can be understood
by the suppression of the wave functions of the KK modes close to the
Planck brane. If there was no suppression of the wave functions, then the
couplings would be suppressed by $M_{Pl}$. The small values of the wave 
functions yield an additional suppression, thus resulting in this
highly suppressed interaction strength.

In fact, this wavefunction suppression at the Planck brane 
is required to maintain the large hierarchy at the loop level 
in the effective 4D theory. Consider for example a scalar field 
living on the Planck wall with mass of ${\cal O}(M_{Pl})$, 
 which has  
renormalizable couplings to scalar 
fields in the bulk. In the 4D effective theory, the bulk 
scalar fields appear as a tower of KK modes, with masses starting  
at ${\cal O}($TeV$)$, whereas the Planck scalar has 
${\cal O}(M_{Pl})$ mass. Without any wavefunction suppression of the 
bulk scalars at the Planck wall, loops of this Planck scalar field 
generate $ {\cal O}(M_{Pl}^2)$ quadratic divergences, which 
would destabilize the hierarchy. With a wavefunction 
suppression 
of $\Omega_0$ , however, the quadratic divergences are ${\cal O}($TeV$)^2$ 
and thus do not destabilize the hierarchy.

\end{document}